\documentclass[twocolumn]{aastex631}

\newcommand{\kms}{km~s$^{-1}$\,}

\newcommand{\msun}{${\cal M}_\odot$\,}

\begin{document}

\renewcommand{\topfraction}{1.0}
\renewcommand{\bottomfraction}{1.0}
\renewcommand{\textfraction}{0.0}

\shorttitle{Visual Orbits}

\title{Orbits  of  Binary  Stars:  from  Visual  Measures  to  Speckle
  Interferometry}

\author{Andrei Tokovinin}
\affil{Cerro Tololo Inter-American Observatory | NFSs NOIRLab Casilla 603, La 
Serena, Chile}
\email{andrei.tokovinin@noirlab.edu}

\begin{abstract}
Knowledge of the orbits of visual  binary stars has always been one of
the  fundamentals  of astronomy.   Based  historically  on the  visual
measures,  nowadays  the orbits  rely  more  (or exclusively)  on  the
accurate speckle data. This prompts  reconsideration of the methods of
orbit  calculation, undertaken  here and  illustrated by  20 examples,
from  accurate  to drastically  revised  and  tentative orbits.   Good
understanding  and critical  assessment of  the  input data  is a  key
requirement,  especially concerning  visual  measures. Combination  of
visual and speckle data is  still needed for long-period binaries, but
the relative  weights must  match their  respective errors.   When the
orbit can be fully constrained only  by accurate speckle data, the old
measures  should be  ignored.   Orbits can  be  classified into  three
grades: A  --- fully  constrained, B ---  semi-constrained, and  C ---
preliminary  or tentative.   Typical use  cases of  visual orbits  are
listed.  Accurate parallaxes from Gaia, together with the orbits, will
greatly  expand  the  data   on  stellar  masses.   Continued  speckle
monitoring  will be  an essential  complement  to Gaia,  but the  vast
amount of  new pairs will restrict  future work on orbits  to the most
interesting or relevant objects.
\end{abstract} 
\keywords{binaries:visual}

\section{Introduction}
\label{sec:intro}

Orbital  motion   of  resolved  binary  stars,   by  tradition  called
``visual'', is  evidenced by change  of their relative  positions over
time.  When the measurements cover a substantial fraction of the orbit,
its seven  elements can  be determined. The orbit  calculation has  been a
classical  problem  in  astronomy  for over  two  centuries,  and  its
solution    is    well    covered     in    papers    and    textbooks
\citep{Aitken1935,Heintz1978}.  At    first  glance,  this   is  a
classical   data  modeling   problem   where   standard  methods   are
applicable.  However,  several  aspects present  specific  challenges,
namely   the  non-linear   relation  between   data  and   parameters,
an insufficient coverage,  and unreliable or ambiguous  measurements. The
methods  of orbit  computing have  evolved  over time,  driven by  the
increasing computing power and better  data, so most textbook methods
have nowadays  become obsolete.  New methods  emerge, mostly  based on
statistical approaches and tailored  to particular needs, e.g.  orbits
of directly  imaged exoplanets \citep{Stojanovski2024} or  a dynamical
analysis   involving  radial   velocities   (RVs)  and   accelerations
\citep{Brandt2021}.

In this work, I share my  experience of computing orbits of relatively
close  resolved  pairs using  both  historic  visual measurements  and
modern speckle interferometry,  and formulate several recommendations.
The current  approach to orbit  fitting, described by  \citet{VB6}, is
revised here  because the quality  and quantity  of the input  data is
steadily  improving  owing  to   the  ongoing  speckle  interferometry
programs   and  other   techniques   such  as   adaptive  optics   and
long-baseline  interferometers.   The  role  of the  old  visual  data
diminishes correspondingly.  Typical  ``visual'' binaries are nowadays
closer and move  faster than in the epoch of  visual measures.  In the
past, partial  coverage of a  long-period orbit was a  major obstacle,
and methods  of computing orbits  from short observed arcs  were given
particular attention.  Now  we often face the  opposite situation when
the orbit is not known even after covering several periods because the
existing  measurements are  not  frequent enough.   A similar  problem
occurs in  fitting astrometric  and spectroscopic  orbits to  the Gaia
data \citep{EDR3} because  the temporal sampling is  determined by the
satellite's  scanning  law, and  nothing  can  be  done to  modify  it
\citep{Halbwachs2023,Holl2023}.

The quality and  reliability of the input data is  the main reason why
the  orbit  computing is  so  special;  it precludes  automatic  orbit
determination, which would otherwise  appear attractive.  The relative
positions measured visually should be treated as estimates rather than
as real  measurements with known errors.   Modern speckle measurements
also  need   a  critical  assessment.   Piecing   together  incomplete
information  makes calculation  of  some visual  orbits a  challenging
task.   Unreliable   or  insufficient   data  often  lead   to  poorly
constrained or plainly wrong orbits.

The input data used for orbit determination and the methods of fitting
Keplerian orbits  are reviewed  briefly in  Section~\ref{sec:calc}.  A
simplified system of three grades  is proposed, replacing the obsolete
5-grade classification  tailored to the  quality of the old  data. The
science  resulting   from  visual   orbits  is     outlined  in
Section~\ref{sec:use}.      Section~\ref{sec:examples}     illustrates
important  aspects of  orbit  calculation by  examples  taken from  my
current  work.   Section~\ref{sec:rec} formulates  recommendations
based on my experience and discusses future trends.

\section{Orbit Calculation}
\label{sec:calc}

\begin{figure}[ht]
\epsscale{1.0}
\plotone{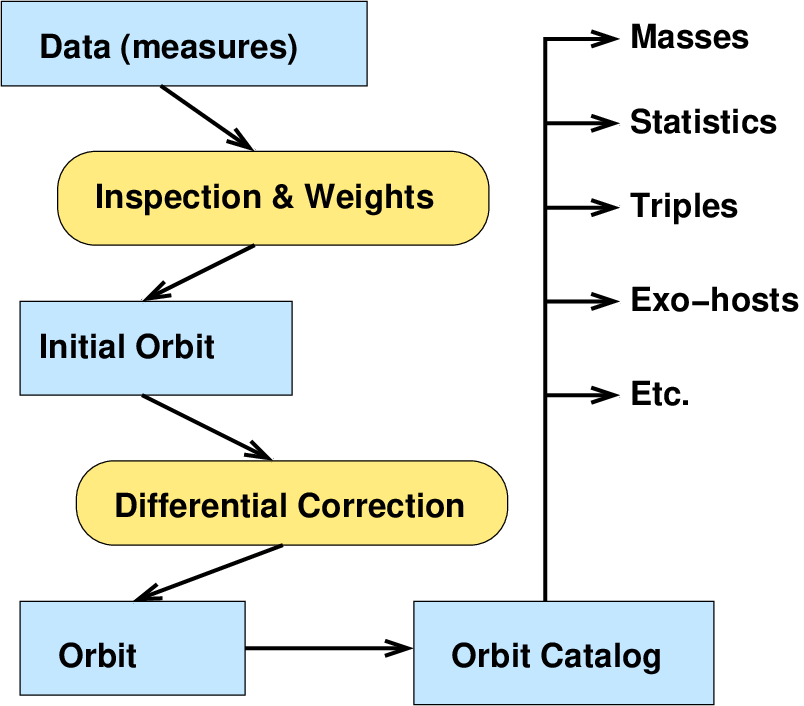}
\caption{Flow chart of orbit calculation.
\label{fig:algorithm} }
\end{figure}

The methods of  orbit calculation have evolved over  two centuries.  A
multitude  of  historic  orbit-computing  methods  \citep{Aitken1935},
invented to alleviate  the burden of calculus,  nowadays are obsolete.
Here, the  modern approach  to orbit  calculation, illustrated  by the
flow chart in Figure~\ref{fig:algorithm}, is outlined.

\subsection{Position Measurements}
\label{sec:data}

When  the  input  data   (position  measurements,  also  called  {\em
  measures}) are  reliable and sufficient, their  modeling (fitting an
orbit)  presents  no  difficulty  and can  be  handled  without  human
supervision.    However,   these   conditions  are   rarely   met   in
practice. Understanding  the input  data and  their limitations  is of
critical importance to orbit calculation. For this reason, most orbits
were and still are computed by  the same people who produce the data.
Orbit  calculation  and  improvement  is the  primary  motivation  for
monitoring the motion of visual binary stars.

Most double-star  measures were  made in the  19th and  20th centuries
visually.   This  technique  is  sometimes  called  filar  micrometer.
Visual measures of close pairs rely on the real-time image analysis by
human     brain     that      resembles     speckle     interferometry
\citep{Couteau1978,Heintz1978}.    These     measures    are    always
subjective. Typically, separations are  estimated with larger relative
errors  than  position angles.   After  correction  of the  systematic
over-estimation of  separations, the  measurements by  the experienced
visual  observer,   P.~Couteau,  show   random  errors   of  0\farcs03
\citep{Tok1983}. The  errors of  visual measures  are not  provided by
their authors,  and the fraction  of discrepant measures  and spurious
visual resolutions is substantial.

\citet{VB6} present a thorough statistical evaluation of the errors of
visual measures,  estimated a posteriori  from their residuals  to the
orbits.   They  parametrize  the   errors  by  the  diffraction  limit
$\lambda/D$ of the telescope used ($\lambda$ is the wavelength and $D$
is  the  aperture  diameter),   with  coefficients  depending  on  the
experience of each observer and the number of averaged measures. Their
Figure~4  indicates  typical  residuals   of  0\farcs07  and  relative
residuals in separation of $\sigma_\rho /\rho \sim 0.12$, with a large
scatter and no  obvious dependence on the orbit  quality.  Orbits were
typically based on 20--100 such measures (or, better said, estimates).

Almost  two centuries  of  visual double-star  observations created  a
legacy data set assembled in  the Washington Double Star (WDS) catalog
\citep{WDS}.\footnote{  \url{https://crf.usno.navy.mil/wds/}}  We  are
indebted to the observers  of the past and use their  data as a ``time
machine'' that  traces orbital motions  back in time.   Nowadays, most
measurements  are  made  using  speckle  interferometry  at  large  or
moderate-size     telescopes.      Other    methods     (long-baseline
interferometers, adaptive optics,  ground- and space-based astrometry)
are also relevant.  Their review is outside the scope of this paper.

Here  I  use mostly  speckle  measures  made  at  the 4.1  m  Southern
Astrophysical  Research (SOAR)  telescope using  the HRCam  instrument
\citep{TMH10,HRCam};  the   latest  papers   \citep{SAM22,SAM23}  give
references to prior publications of the  results.  As of May 2024, the
HRCam database  contained 27,184  measurements of  relative positions.
The pixel  scale and  orientation are  calibrated by  observing slowly
moving  binaries  with separations  on  the  order of  1\arcsec;  this
calibrator  set  was referenced  to  Gaia  in \citet{SAM21},  and  the
recommended  minor  corrections  to  the  measures  made  before  2021
(subtract 0\fdg2 from  the position angles and  divide the separations
by 1.0053) are applied here. For most binaries, residuals of the HRCam
positions  from the  orbits  are  within 2\,mas,  so  such errors  are
assumed.   The less  accurate speckle  measures at  4 m  telescopes by
CHARA and other teams are given  here errors of 5\,mas.  Note that the
weighting  system  adopted by  \citet{VB6}  accounts  for the  smaller
speckle errors  only by the  aperture-size factor $\lambda/D$.   So, a
typical  70\,mas  error  of  visual measures  with  0.7  m  telescopes
corresponds to 12\,mas  with 4 m telescopes, and  this weighing scheme
severely  under-estimates  the  real  accuracy  of  the  speckle  data
relative to the visual measures.

When  the  magnitude  difference  $\Delta   m$  is  small  (i.e.   the
components'  fluxes  are  comparable),   the  position  angle  can  be
``flipped'' by 180\degr.  The  classical speckle interferometry yields
angles modulo 180\degr  ~for all binaries, although  this ambiguity can
be removed  using advanced techniques  such as image  restoration when
$\Delta  m$  is  not  too small.  The  quadrant  ambiguity  sometimes
presents substantial problems, as illustrated by the examples below.

\subsection{Orbital Elements}
\label{sec:el}

Motion of a resolved binary system is described by the seven classical
Campbell orbital elements: period $P$, time of periastron passage $T$,
eccentricity $e$, semimajor axis $a$  in angular units, position angle
of the ascending  node $\Omega$, longitude of the  periastron from the
ascending    node    $\omega$,    and    orbital    inclination    $i$
\citep{Aitken1935,Heintz1978,Brandt2021}.  If  radial velocities (RVs)
of the primary and secondary components are measured, their amplitudes
$K_1$ and $K_2$ and the  systemic velocity $\gamma$ can be determined,
increasing the total number of orbital elements to ten.

For  pairs  with  direct (counterclockwise)  motion,  the  inclination
$i<90\degr$,  for the  retrograde (clockwise)  motion $90\degr  < i  <
180\degr$.   Face-on orbits  with  $i \approx  0\degr$  or $i  \approx
180\degr$ are  degenerate because  the elements $\Omega$  and $\omega$
are  strongly  correlated  (both  define orientation  of  the  orbital
ellipse on the sky). In such case, it is appropriate to set $\omega=0$
and  to fit  the  remaining  five elements.  Circular  orbits with  $e
\approx 0$ are also degenerate because the periastron is arbitrary; by
setting $\omega=0$,  we effectively  redefine the  element $T$  as the
time  of the  nodal passage.   In a  circular face-on  orbit, $T$  and
$\Omega$ also  become degenerate,  and only three  elements $P,  T, a$
suffice to  describe the motion  (in such case,  $T$ is the  time when
$\theta = 0$).

\subsection{Orbit Determination}
\label{sec:fit}

Let $x_k(t_i)$ be  the measurements, e.g. the  position angle $\theta$
and separation $\rho$ at time  $t_i$ or, alternatively, the offsets of
the  companion in  declination and  R.A.  ($k=1,2$).   The measurement
errors are $\sigma_{k,i}$.   The vector of the  seven orbital elements
${\mathbf  p}$ (parameters)   determines  the ephemeris  positions
$\hat{x}_k(t_i, {\mathbf  p})$.  Deviations  of measurements  from the
orbit (residuals) are characterized by the standard $\chi^2$ metric,
\begin{equation}
\chi^2 = \sum_{i,k} \frac{ [x_k(t_i) - \hat{x}_k(t_i, {\mathbf
      p})]^2}{\sigma_{k,i}^2} .
\label{eq:chi2}
\end{equation}
Fitting  an  orbit  to  measurements implies  finding  the  parameters
${\mathbf  p}$  that minimize  $\chi^2$  ---  a typical  data-modeling
problem.  This approach can be  extended by including additional types
of   data,   such    as   RVs   \citep{Pourbaix2000,Lucy2018}   and/or
accelerations \citep{Brandt2021}; however, additional parameters (e.g.
masses, distances,  or RV  amplitudes) must be  fitted in  such cases.
This  formulation  assumes  implicitly  that  the  measurement  errors
$\sigma_{k,i}$ are known and normally  distributed, which is not quite
true for  position measurements, especially those  made visually. 
  Nevertheless,  this  approach  with   suitably  assigned  errors  is
  universally adopted.   Evaluation of  the errors (which  are either
not published together with the measures  or are only lower limits) is
the cornerstone of  the orbit computing: different  orbits are derived
from the same data by adopting different errors.

The specifics of fitting a visual  orbit is the high dimensionality of
the parameter  space, the  diversity of  situations, depending  on the
orbit  orientation  and  its  coverage, and  the  non-linear  relation
between  parameters and  data.   The latter  aspect  can be  partially
improved by replacing the Campbell  elements $a, \Omega, \omega, i$ by
their  combinations  $A,  B,  F, G$,  called  the  Thiele-Innes  (T-H)
elements \citep{Aitken1935, Heintz1978}.   Linear relation between the
T-H elements and the measurements allows us to solve the least-squares
(LS)  problem  by  the  standard  matrix  method,  but  the  remaining
non-linear elements  $P, T, e$ need  to be known to  take advantage of
this parametrization.

The high dimensionality of the  parameter space and the non-linearity,
combined with insufficient data, often result in a complex topology of
the $\chi^2$  surface with multiple  local minima.  The shape  of this
hypersurface also depends on the  adopted errors $\sigma_{i,k}$ and is
strongly influenced by the highly deviant or wrong measurements. These
aspects  make the  orbit  fitting  a tricky  problem.  If the  orbital
elements are known  approximately (e.g. from prior work),  they can be
corrected (improved) via linearization and LS fitting.  So, finding an
initial approximation of an orbit is the first and critical step.

\subsection{Initial Approximation}
\label{sec:init}

An obvious approach to finding the global $\chi^2$ minimum is to split
the parameter space  into the linear and non-linear  domains.  This is
implemented in  the {\em  grid search} method  \citep{Hartkopf1989}. A
suitably  fine  grid in  the  3-dimensional  space of  the  non-linear
parameters $P,T,e$  is defined and  explored by brute force.   At each
grid point, the  T-H elements are fitted, and the  grid point with the
minimum $\chi^2$ is  taken as the initial approximation  for the final
LS refinement of all elements.

The  grid  search  method  is perfect  for  reliable  data.   However,
rejection (or down-weighting) of poor data is a critical step in orbit
fitting.  Furthermore, flips of the  position angles by 180\degr, when
allowed, critically affect  the orbit.  So, instead of  relying on the
blind  grid search,  I examine  the input  data and  find the  initial
approximation heuristically.

The third Kepler  law gives a crude estimate  of the orbital period
$P^*$ (in years) as
\begin{equation}
P^* = (\rho/\varpi)^{3/2} M^{-1/2} ,
\label{eq:Pstar}
\end{equation}
where $\rho$ is the typical  separation, identified with the semimajor
axis $a$, $\varpi$ is  the parallax, and $M$ is the  mass sum in solar
units. The parallaxes are nowadays  known from Gaia or, otherwise, can
be  estimated photometrically  assuming main-sequence  stars; $M  = 2$
\msun  is a  good  starting assumption.  Statistically, the  condition
$\rho \approx a$  holds within a factor of 2,  and the crude estimates
$P^*$ typically differ from the true  periods within a factor of $\sim
3$.

I use the  interactive IDL code {\tt orbit} \citep{orbit}  to plot the
measurements and  to find the  initial approximation.  At  this stage,
the outliers  are rejected and  the quadrant flips are  adjusted.  The
sense  of  the  orbital  motion suggests  to  start  with  $i=60\degr$
(counterclockwise)  or  $i=120\degr$  (clockwise);  the  time  of  the
closest  approach gives  an  idea  of the  periastron  epoch $T$,  the
position angle at maximum  separation gives $\Omega$, assuming $\omega
\approx 180\degr$, $P  = P^*$, $a= \rho$, and $e  =0.5$.  A few manual
tweaks  of the  crude  starting  elements usually  suffice  to get  an
initial  orbit   that  resembles  the  data   approximately.   If  the
separation remains almost constant, we can start with a circular orbit
by setting  $e=\omega=0$, and allow  a non-zero eccentricity  later in
the LS fit.  If the LS solution  converges to a negative $e$, make the
following replacements: $T \rightarrow T \pm P/2$, $\omega \rightarrow
\omega \pm 180\degr$, $e \rightarrow -e$.

It is often helpful to reduce  the dimension of the parameter space by
fixing most  elements and fitting  only one or  two. In this  way, the
initial orbit can  be made closer to the data.  Experience helps us to
decide which  elements need to be  fitted or changed manually  at this
initial stage to  get a reasonable first  approximation. Sometimes the
data     can    match     several    radically     different    orbits
(Section~\ref{sec:drastic}).

\subsection{Orbit Correction}
\label{sec:corr}

The final  adjustment of  the orbital  elements  is done  by the  standard
iterative LS method, linearizing the model in the
vicinity of the initial parameters ${\mathbf p}_0$:
\begin{equation}
\hat{x}_k(t_i, {\mathbf  p}) \approx \hat{x}_k(t_i, {\mathbf  p}_0) +
\sum_j \frac{\partial \hat{x}_k(t_i, {\mathbf p})}{\partial p_j} (p_j - p_{0,j}) ,
\label{eq:lin}
\end{equation}
where  the  index  $j$  denotes the  fitted  parameters.  The  partial
derivatives of the  Keplerian orbit can be  calculated analytically or
numerically  (the  latter  approach  is used  in  {\tt  orbit},  while
analytic gradients  were programmed  in its earlier  FORTRAN version).
The  linearization allows  minimization  of $\chi^2$  by the  standard
Levenberg-Marquardt method  which combines  the steepest  descent with
the full LS solution \citep{NumRec}.   The formal errors of the fitted
elements and their covariances are  determined in the process, as well
as  the goodness  of  fit metric  $\chi^2$.  If  the  data errors  are
assigned correctly, we  expect to obtain $\chi^2/(N-J)  \approx 1$ for
$N$ measurements  and $J$ fitted  parameters. This metric  is computed
separately  for  each  type  of data  (e.g.  separations  and  angles)
specified by the index $k$.

As  noted  above,  in  some  cases the  set  of  orbital  elements  is
degenerate (the matrix  of the corresponding LS  problem is singular),
so some  elements (or  their combinations) must  be fixed.   Often the
data do not constrain the elements  well enough.  For example, a short
observed arc does not constrain the  period $P$ and the semimajor axis
$a$.   However, these  elements  are usually  strongly correlated,  so
their ratio  $a^3/P^2$ that defines  the mass sum is  constrained much
better than  $P$ and $a$ \citep{Lucy2014}.   Another typical situation
is  when the  binary is  not resolved  at close  separations near  the
periastron of  an eccentric orbit,  so the data  match a  range of
eccentricities $e$.  If the parallax is known, calculation of the mass
sum by  the third Kepler's  law and  its comparison with  the expected
masses  helps us  to  select values  of  poorly constrained  elements.
Generally  speaking, the  mass  sum is  a good  sanity  check for  all
orbital solutions.

Typically,  new  measurements are  used  to  improve known  orbits  by
differential  correction.   However, if  the  initial  (old) orbit  is
wrong, the  differential correction  will not change  it dramatically,
and the updated orbit will remain wrong  as well. In such cases, a new
initial orbit is needed, and the orbit calculation must restarted from
scratch.

\subsection{Orbit Catalog and Grades}
\label{sec:cat}

Data on all visual orbits are assembled in the electronic catalog ORB6
maintained at USNO\footnote{\url{https://crf.usno.navy.mil/wds-orb6/}}
and  described  by \citet{VB6}.   The  quality  of visual  orbits  was
traditionally  estimated by  experts, with  grade 1  for the  best and
grade 5 for the worst  (tentative) orbits.  The subjective grades were
replaced by the automatic grading system developed by \citet{VB6}.  At
that time,  most orbits were  still based on  the visual data,  so the
grading algorithm was tailored to  this reality and ``trained'' on the
subjective  grades assigned  by  the experts.   For  example, a  large
number of measurements was required  for grade 1 orbits, although only
$\sim$10 accurate  positions with  a good  phase coverage  suffice for
computing an accurate orbit.   A simple one-dimensional grading scheme
cannot capture  the diversity of situations  and the multi-dimensional
nature of the orbital parameters.   What really matters are the errors
(i.e.  the confidence  intervals) of the elements  or, more generally,
the  constraints provided  by  the  data.  I  propose  to replace  the
existing orbit grades by a simple 3-tier classification:
\begin{itemize}
\item
{\it Grade A}  --- fully constrained orbits  with meaningful estimates
of the errors (or confidence intervals) for all elements.  The orbit quality is characterized by the errors better than by the discrete grades.

\item
{\it Grade B} --- semi-constrained  orbits where the error of the
period  is  determined,  but  one or  more  other  poorly  constrained
elements can be fixed in the orbit fitting.

\item
{\it  Grade   C}  ---   unconstrained  or  preliminary   orbits  (this
corresponds  roughly  to the  old  grades  4  and 5).   No  meaningful
estimates of the elements' errors can be given.

\end{itemize}

The low reliability of visual  measures has been generally recognized,
and traditionally the observers were  not blamed for producing deviant
data points or spurious resolutions.  The visual orbits based on these
data inherited the  culture of error tolerance  which differs markedly
from  the  more  strict  attitude to  spectroscopic  orbits.   Another
justification  for  publishing  low-quality  or  tentative  orbits  is
related to the long orbital periods.  A tentative orbit estimated from
the short observed  arc may be completely off, but  it is still better
than nothing,  with little  prospect of  improvement in  the following
decades owing to the slow motion.  In the ORB6 catalog (queried on 2024 August
1),  1816  orbits (59\%)  are  of  grades  4  and 5  (preliminary  and
tentative), and  only 1340 orbits  of grades 1, 2,  and 3 are  more or
less reliable.

In  his paper  ``Is  this orbit  really necessary?'',  \citet{vdB1962}
questioned the  established practice of publishing  poorly constrained
low-grade orbits. He  also argued that well-defined  orbits should not
be revised in response to new  measures. However, the high accuracy of
speckle interferometry calls  for the correction of even  the best old
orbits which were  based on the less accurate historic  data.  I agree
that  such corrections  should not  be  too frequent,  they should  be
justified by the well-established systematic residuals to the existing
orbits. Two such examples are given below in Section~\ref{sec:good}.

Publication  of  orbits   poorly  constrained  by  the  currently
available data  can be justified  by their subsequent  use.  Tentative
orbits represent  the existing  measures and  help in  planning future
observations  (e.g.  predict  the  critical  phases near  periastron).
They serve  for computing   past  and future  positions on  the sky
(improve the  coordinates by accounting  for the orbital  motion), and
are valuable for the study  of stellar systems that contain additional
bodies (e.g.  inner subsystems, disks,  or exoplanets).  A high demand
for  evaluating  orbital  parameters  even from  short  arcs  prompted
the development of new methods and software \citep{Blunt2020,Brandt2021}.

Owing to  the ongoing  speckle programs,  the content  of the  ORB6 in
terms of the number of orbits and their quality is steadily improving.
We should keep  in mind, however, that the knowledge  of visual orbits
does  not advance  the  science  by itself,  but  rather  serves as  a
fundamental   basis    for   further   research,   as    outlined   in
the following Section.

\section{Use of Visual Orbits}
\label{sec:use}

Discovery of double stars by  visual resolution, their monitoring, and
calculation  of  orbits  based  on the  accumulated  measures  is  the
classical pattern.   This research program  has been executed  for two
centuries  by several  generations  of  astronomers. Its  continuation
(compute   new  orbits   and  improve   the  existing   ones)  appears
natural. Without orbits, the past measures remain essentially useless,
so monitoring  of ``everything  that moves'' seems  to be  an adequate
tribute  to   our  predecessors,  extending  their   effort  into  the
future. However,  the use  of the telescope  time and  other resources
cannot be justified  only by the need to improve  the ORB6 catalog: by
itself, it  does not produce  new science.  The current use  of visual
orbits is reviewed below.

\subsection{Measurement of Masses}
\label{sec:mass}

Historically,  visual  orbits served  to  measure  stellar masses,  to
establish   the  mass-luminosity   relation,  and   to  test   stellar
evolutionary models  \citep{Andersen1991}. For the latter,  a relative
accuracy  of the  mass on  the order  of 2\%  or better  is considered
adequate.  The  mass sum  of a  visual binary $M$  is computed  by the
third Kepler law expressed by  equation \ref{eq:Pstar}, where $P^*$ and
$\rho$ are replaced by the true period $P$ and the semimajor axis $a$,
respectively:
\begin{equation}
M = (a/\varpi)^{3} P^{-2} .
\label{eq:mass}
\end{equation}
Only  a  minority of  very  good  orbits  in  ORB6 have  the  quantity
$a^3/P^2$   precise   enough   for  useful   tests   of   evolutionary
models. Moreover, the parallax $\varpi$  must be known with a matching
accuracy of  0.6\%, and here  we have a  problem. About two  thirds of
objects   in  ORB6   do  not   have   any  astrometry   in  Gaia   DR3
\citep{Chulkov2022}, while  the Hipparcos parallaxes are  not accurate
enough. For the remaining orbital  binaries, the Gaia DR3 does
not  account  for  the  orbital  motion  while  fitting  the  standard
5-parameter astrometric  solutions, so  the parallaxes of  most visual
binaries  are biased.  The bias  is corrected  only for  binaries with
astrometric  orbits in  Gaia;  their  periods are  shorter  than 3  yr
\citep{Halbwachs2023}.

There are several  ways to circumvent the  Gaia limitations concerning
visual  binaries. When  a visual  binary is  accompanied by  a distant
bound tertiary companion,  its accurate Gaia parallax can  be used for
measuring    the   masses    (see    for    example   00024+1047    in
Section~\ref{sec:drastic}).    Alternatively,   the   Gaia   data   on
individual transits, which will become public in the next data release
DR4, can be used in combination  with the speckle measures to fit both
the   astrometry   and   the    orbit   jointly.   The   ORVARA   code
\citep{Brandt2021} is  a suitable  tool for  doing this.  Finally, the
combination of  visual and  double-lined spectroscopic orbits yields  distances and
masses   directly  without   the   need   of  astrometric   parallaxes
\citep{Pourbaix2000}; however,  a very high accuracy  of both resolved
measures and RVs is mandatory for accurate mass measurement.

The  majority of  visual binaries  with known  orbits have  solar-type
components, and masses of such stars  are already known quite well, so
additional  similar  data have  little  value.   The current  interest
focuses  on less  explored stars,  for  example on  the M-type  dwarfs
\citep{Mann2019,Vrijmoet2022},   brown   dwarfs   \citep{Rickman2024},
pre-main  sequence stars  \citep{Rizzuto2020}, or  very massive  stars
\citep{Schaefer2016,Klement2024}.   In  these   cases,  even  modestly
accurate masses contribute new knowledge.


\subsection{Statistics}
\label{sec:stat}

A catalog of  orbits can be exploited for statistical  studies. One of
the classical topics is  the period-eccentricity relation. Analysis of
spectroscopic and visual orbits established a trend of increasing mean
eccentricity             with            increasing             period
\citep{Finsen1936,TokKiy2016}. Using  the Gaia data,  the eccentricity
distribution  of  very  wide  binaries  was  evaluated  statistically,
confirming   and   extending   this   trend   to   large   separations
\citep{Hwang2022}.

The  major  obstacle   in  using  ORB6  for  the   statistics  is  its
heterogeneous  and random  content.  The  calculation of  an orbit  is
driven  be the  data availability,  which  varies greatly,  as can  be
inferred from the  examples in this paper.  Very  eccentric orbits are
particularly demanding in this respect, and they are under-represented
in  the catalogs  \citep[so-called Finsen's  effect,][]{Finsen1936}.  A
large fraction of low-grade orbits  and their drastic revisions reduce
the  confidence  in the  distributions  of  periods or  eccentricities
derived from  ORB6. To  overcome these caveats,  one can  restrict the
input  samples,   for  example  to  solar-type   stars  within  25\,pc
\citep{Raghavan2010}      or     to      nearby     M-type      dwarfs
\citep{Vrijmoet2022}. In  these smaller  samples, the  completeness of
known visual orbits can be  tested, and additional observations can be
planned to improve it.

The orbital  angles $\Omega, \omega,  i$ are not interesting  in their
own right because  binary orbits are oriented  randomly.  However, the
orbital inclinations serve to probe  relative alignment of orbits with
stellar spins  \citep{Weis1974}, with orbits of  transiting exoplanets
\citep{Lester2023}, or  with inner  subsystems in  stellar hierarchies
\citep{Tok2017,Tok2021a}.   Although the  content  of  ORB6 is  highly
heterogeneous, these studies postulate  that the calculation of visual
orbits  is  not influenced  by  the  orientation  of stellar  axes  or
planetary orbits, hence any revealed correlations are genuine.

\subsection{Hierarchical Systems}
\label{sec:mult}

Hierarchical  systems   of  three  or   more  stars  have   a  diverse
architecture   related  to   their  formation   and  early   evolution
\citep{review,Offner2023}.   Relative orientation  of inner  and outer
orbits  in  triple  systems,   period  ratios,  mutual  dynamics,  and
potential resonances  present a rich research  field where calculation
of orbital elements plays an essential role \citep{Borkovits2016}.

As  an  example,  consider  HIP   12548,  a  classical  visual  binary
discovered  by  R.    Aitken  in  1931  and   composed  of  solar-type
stars. Historic measures, complemented  by the speckle interferometry,
define its 106 yr orbit  quite well. However, spectroscopic monitoring
revealed that  each visual component has  a variable RV, so  this is a
2+2  quadruple  system  with  inner   periods  of  5.07  and  0.30  yr
\citep{CHIRON-8}.  The  5 yr subsystem produces  measurable deviations
(wobble)  in  the  positions,   allowing  us to  establish  the  relative
orientation of  the subsystem with  respect to the outer  orbit.  This
case  resembles the  classical  visual  binary $\alpha$~Gem  (Castor),
where each visual  component is a close  spectroscopic pair. Recently,
both spectroscopic subsystems in Castor were spatially resolved by the
CHARA interferometer,  showing that the  inner orbits are  not aligned
with the  outer visual orbit  \citep{Torres2022}.  In both  cases, the
classical visual  binaries are  revealed as hierarchical  systems, and
their orbits  acquire new  significance.  Needless  to say  that using
such visual pairs for testing stellar evolution makes no sense without
measuring  masses  of all  components,  because  they are  not  simple
binaries.

\subsection{Objects of Special Interest}
\label{sec:obj}
 

Calculation of visual orbits is  required for various reasons when the
stars present  some special interest  or significance. A  typical case
are exohosts,  where the knowledge  of orbits constraints the  size of
truncated  protoplanetary disks  and  thus informs  us  on the  planet
formation mechanisms or, statistically,  probes the relative alignment
between   transiting   planets  and   the   orbits   of  their   hosts
\citep{Lester2023}.  The orbital motion  of directly imaged planets is
being monitored,  and even  short observed arcs  help to  evaluate the
range of possible orbital parameters \citep{Stojanovski2024}.

The pre-main  sequence (PMS) star  HD~98800 is a 2+2  quadruple system
where  the  outer  pair  A,B   is  a  classical  visual  binary  I~507
(11221$-$2447).  Recent interferometric  observations have established
the orientation  of the inner  subsystems relative to the  outer orbit
\citep{Zuniga-Fernandez2021}.  The inner pair Ba,Bb is surrounded by a
debris disk which is nearly perpendicular  to its orbit.  The A,B pair
is closing down, and during the upcoming conjunction in 2026--2027 the
disk  will occult  the subsystem  Aa,Ab, offering  a unique  chance to
study the  detailed disk  structure by  photometric monitoring  of the
occultation.   The  knowledge  of  the outer  orbit  is  critical  for
interpretation of these data, and  the historic visual measures of A,B
play an important role here.

Yet  another emblematic  case is  the Cepheid  Polaris ($\alpha$~UMi),
member of  a triple  system. The  orbit of the  inner 30  yr subsystem
determined using  RVs and  resolved measures  yields the  Cepheid mass
\citep{Evans2024}.   Owing  to  the large  magnitude  difference,  the
position  measures  come   from  telescopes  in  space   or  from  the
long-baseline  interferometry; this  ``visual'' binary  is beyond  the
reach of classical visual observers of the past.

\section{Examples of Orbits}
\label{sec:examples}

The  specifics of  the  visual orbit  calculation  and the  associated
caveats  are illustrated  here  with 20  orbits  recently computed  or
revised  by  the  author.   We   start  with  two  accurate  and  very
well-constrained  orbits  based  on  both  visual  and  speckle  data,
followed by several examples of orbits relying entirely on the speckle
measurements.  Then two interesting cases illustrating spurious visual
resolutions of very  close pairs are presented.   Examples of dramatic
revisions of  published orbits and  a few tentative  first-time orbits
close this Section.  The orbital  elements and their errors are listed
in Table~\ref{tab:orb}.   Table~\ref{tab:obs}, available in  full only
electronically, contains  all measurements, their adopted  errors, and
the residuals  to the orbits.  According  to the rules adopted  in the
Washington  Double  Star  Catalog,  WDS \citep{WDS},  each  binary  is
identified by its WDS code based on the J2000 position and by a unique
string called  ``Discoverer Designation'' (DD); the  DDs, however, are
barely  used  outside  the  double-star  community.   In  hierarchical
systems, a common WDS code refers to several pairs with different DDs.
An  alternative designation  scheme based  on components,  rather than
pairs, is adopted in the Multiple Star Catalog \citep{MSC}.


\startlongtable

\begin{deluxetable*}{l l cccc ccc cc}    
\tabletypesize{\scriptsize}     
\tablecaption{Visual Orbits
\label{tab:orb}          }
\tablewidth{0pt}                                   
\tablehead{                                                                     
\colhead{WDS} & 
\colhead{Discoverer} & 
\colhead{$P$} & 
\colhead{$T$} & 
\colhead{$e$} & 
\colhead{$a$} & 
\colhead{$\Omega$ } & 
\colhead{$\omega$ } & 
\colhead{$\i$ } & 
\colhead{Grade }  &
\colhead{Previous} \\
  &
\colhead{Designation}  & 
\colhead{(yr)} &
\colhead{(yr)} & &
\colhead{(arcsec)} & 
\colhead{(deg)} & 
\colhead{(deg)} & 
\colhead{(deg)} &   &
\colhead{Orbit\tablenote{a}} 
}
\startdata
00024$+$1047 & A 1249 AB & 58.86 & 2002.91 & 0.806 & 0.2004 & 66.4 & 18.6 & 110.6 & A & Zir2003 \\
             &     & $\pm$1.25 & $\pm$0.55 & $\pm$0.022 & $\pm$0.0031 & $\pm$2.2 & $\pm$5.9 & $\pm$1.4&     &  \\
04123$+$0939 & STT 74 & 110.09 & 1996.83 & 0.905 & 0.3332 & 111.9 & 17.0 & 108.8 & A & Msn2019 \\
             &     & $\pm$1.81 & $\pm$0.30 & $\pm$0.006 & $\pm$0.0038 & $\pm$0.9 & $\pm$2.6 & $\pm$0.8&     &  \\
04386$-$0921 & TOK 387 & 4.239 & 2019.351 & 0.450 & 0.0509 & 40.6 & 292.9 & 106.2 & A & \ldots \\
             &     & $\pm$0.035 & $\pm$0.043 & $\pm$0.028 & $\pm$0.0014 & $\pm$1.1 & $\pm$2.7 & $\pm$1.7&     &  \\
05005$+$0506 & STT 93 & 3200 & 1913.563 & 0.80 & 4.108 & 60.6 & 65.5 & 100.3 & C & Izm2019 \\
05251$-$3803 & I 1493 & 250 & 2045.0 & 0.70 & 0.238 & 153.2 & 92.7 & 57.2 & C & \ldots \\
10407$-$0211 & A 1351 & 168.7 & 1992.87 & 0.90 & 0.4742 & 74.2 & 250.1 & 115.2 & C & \ldots \\
13453$+$0903 & BU 115 AB & 800 & 2164.4 & 0.50 & 1.793 & 103.3 & 264.8 & 55.2 & C & Izm2019 \\
14056$-$3916 & I 1575 & 64.9 & 2018.62 & 0.776 & 0.1353 & 119.5 & 290.8 & 160.0 & C & \ldots \\
14160$-$0704 & HU 138 & 142.8 & 1938.26 & 0.571 & 0.4477 & 74.3 & 304.9 & 43.5 & A & Doc1990d \\
             &     & $\pm$2.8 & $\pm$0.69 & $\pm$0.016 & $\pm$0.0068 & $\pm$2.9 & $\pm$1.9 & $\pm$1.9&     &  \\
14453$-$3609 & I 528 AB & 15.933 & 2023.555 & 0.593 & 0.0459 & 236.5 & 66.6 & 25.6 & A & Tok2022f \\
             &     & $\pm$0.162 & $\pm$0.090 & $\pm$0.028 & $\pm$0.0023 & $\pm$18.5 & $\pm$17.3 & $\pm$7.7&     &  \\
15493$+$0503 & A 1126 & 1000 & 2005.83 & 0.90 & 0.291 & 16.2 & 8.6 & 38.6 & C & Gomez2022 \\
16090$-$0939 & WSI 85 & 9.51 & 2020.89 & 0.97 & 0.0986 & 139.4 & 319.2 & 83.8 & B & \ldots \\
             &     & $\pm$0.15 & $\pm$0.30 & fixed & $\pm$0.0273 & $\pm$3.4 & $\pm$19.2 & $\pm$5.2&     &  \\
16103$-$2209 & TOK 860 & 7.72 & 2023.10 & 0.314 & 0.0479 & 121.2 & 111.9 & 95.9 & A & \ldots \\
             &     & $\pm$0.60 & $\pm$0.15 & $\pm$0.078 & $\pm$0.0013 & $\pm$1.5 & $\pm$10.6 & $\pm$1.5&     &  \\
16245$-$3734 & B 868 AB & 1.700 & 2022.711 & 0.957 & 0.0194 & 164.9 & 299.7 & 20.0 & B & \ldots \\
             &     & $\pm$0.014 & $\pm$0.083 & $\pm$0.098 & $\pm$0.0114 & $\pm$696.2 & $\pm$674.2 & fixed&     &  \\
16520$-$3602 & RSS 420 Aa,Ab & 6.99 & 2018.503 & 0.40 & 0.0979 & 18.6 & 103.0 & 139.5 & B & \ldots \\
             &     & $\pm$0.06 & $\pm$0.057 & fixed & $\pm$0.0013 & $\pm$3.6 & $\pm$2.1 & $\pm$1.4&     &  \\
17005$+$0635 & CHR 59 & 13.111 & 2014.45 & 0.725 & 0.1063 & 65.6 & 187.7 & 24.9 & A & Tok2022f \\
             &     & $\pm$0.056 & $\pm$0.16 & $\pm$0.032 & $\pm$0.0025 & $\pm$26.6 & $\pm$29.5 & $\pm$10.9&     &  \\
17093$-$2954 & B 330 & 57.75 & 1998.26 & 0.353 & 0.1780 & 65.2 & 126.5 & 45.6 & A & \ldots \\
             &     & $\pm$1.46 & $\pm$0.75 & $\pm$0.027 & $\pm$0.0085 & $\pm$3.8 & $\pm$5.8 & $\pm$3.7&     &  \\
17304$-$0104 & STF2173 AB & 46.538 & 2008.715 & 0.175 & 0.9697 & 151.63 & 325.56 & 99.33 & A & Msn2023 \\
             &     & $\pm$0.019 & $\pm$0.042 & $\pm$0.001 & $\pm$0.0008 & $\pm$0.04 & $\pm$0.35 & $\pm$0.04&     &  \\
17305$-$1006 & RST3978 & 93.93 & 2008.60 & 0.244 & 0.5737 & 97.0 & 91.0 & 78.3 & A & Tok2015c \\
             &     & $\pm$2.03 & $\pm$0.44 & $\pm$0.019 & $\pm$0.0055 & $\pm$0.2 & $\pm$2.8 & $\pm$0.3&     &  \\
19164$+$1433 & CHR 85 Aa,Ab & 6.938 & 2022.495 & 0.577 & 0.0392 & 37.5 & 148.5 & 147.4 & A & Tok2015c \\
             &     & $\pm$0.023 & $\pm$0.083 & $\pm$0.021 & $\pm$0.0011 & $\pm$8.9 & $\pm$10.8 & $\pm$6.0&     &  \\
\enddata 
\tablenotetext{a}{References to previous orbits: 
Doc1990d = \citet{Doc1990d};
Gomez2022 = \citet{Gomez2022};
Izm2019 = \citet{Izm2019};
Msn2019 = \citet{Msn2019};
Msn2023 = \citet{SAM22}; 
Tok2015c = \citet{TMH15};
Tok2022f = \citet{SAM21};
Zir2003 = \citet{Zir2003}}
\end{deluxetable*}

\begin{deluxetable*}{c  r rrr rr l }
\tabletypesize{\scriptsize}
\tablewidth{0pt}
\tablecaption{Positional Measurements and Residuals (Fragment) \label{tab:obs}}
\tablehead{
\colhead{WDS} & 
\colhead{$T$} &
\colhead{$\theta$} & 
\colhead{$\rho$} &
\colhead{$\sigma$} & 
\colhead{O$-$C$_\theta$} & 
\colhead{O$-$C$_\rho$} &
\colhead{Ref.\tablenotemark{a}} \\
&  
\colhead{(yr)} & 
\colhead{(\degr)} &
\colhead{(\arcsec)} & 
\colhead{(\arcsec)} & 
\colhead{(\degr)} &
\colhead{(\arcsec)} &
}
\startdata
00024+1047 & 1905.5500 & 239.9 & 0.3300 & 0.0500 &   -3.6 & -0.0053 & M \\
00024+1047 & 1916.2100 & 236.2 & 0.3400 & 0.0500 &   -2.7 & -0.0014 & M \\
00024+1047 & 1926.7400 & 227.4 & 0.2800 & 0.0500 &   -6.0 &  0.0063 & M \\
00024+1047 & 1979.7400 & 235.8 & 0.2900 & 0.0500 &   -1.0 & -0.0305 & M \\
00024+1047 & 1987.7532 & 231.3 & 0.2520 & 0.0050 &   -0.5 &  0.0017 & s \\
00024+1047 & 1988.6595 & 231.0 & 0.2400 & 0.0050 &   -0.0 &  0.0006 & s \\
00024+1047 & 1991.2500 & 218.0 & 0.1850 & 0.0500 &  -10.3 & -0.0194 & H \\
\enddata
\tablenotetext{a}{
A: adaptive optics;
G: Gaia;
H: Hipparcos;
M: visual micrometer measurement;
S: speckle interferometry at SOAR;
s: speckle interferometry at other telescopes.}
\tablenotetext{}{(This table is available in its entirety in
  machine-readable form) }
\end{deluxetable*}

\subsection{Accurate Orbits}
\label{sec:good}

\begin{figure}[ht]
\epsscale{1.0}
\plotone{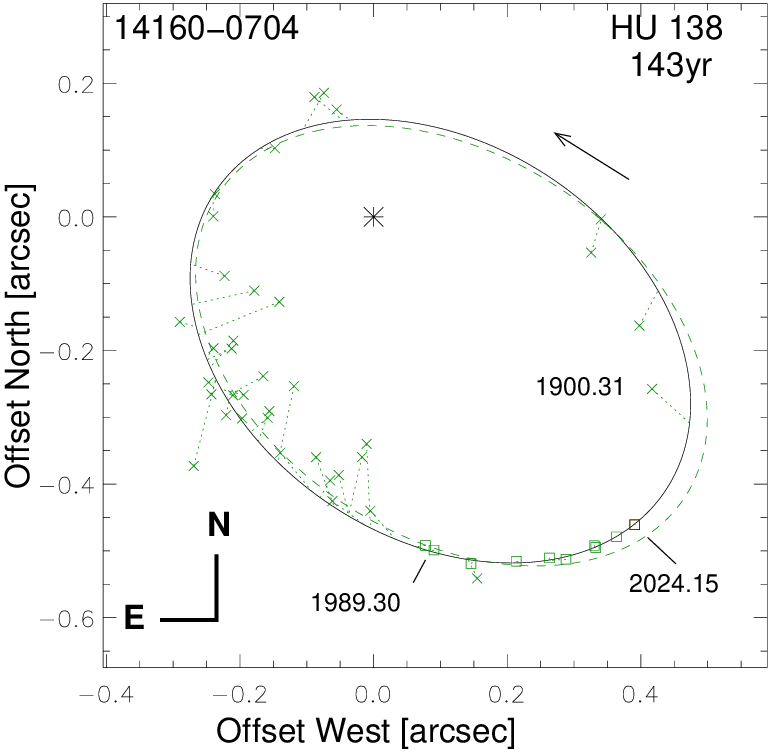}
\caption{The definitive orbit of 14160$-$0704 (HU  138). In this and following
  plots, the primary  component is located  at the coordinate  origin, the
  axis  scale  is  in  arcseconds (North  up,  East  left). The  accurate
  measurements are  plotted as  squares (in  red color  after 2023.0),
  the less accurate (e.g.  visual) ones as crosses. The full line is an orbital
  ellipse, the dashed line plots the previously computed orbit.
\label{fig:accurate1} }
\end{figure}

\begin{figure}[ht]
\epsscale{1.1}
\plotone{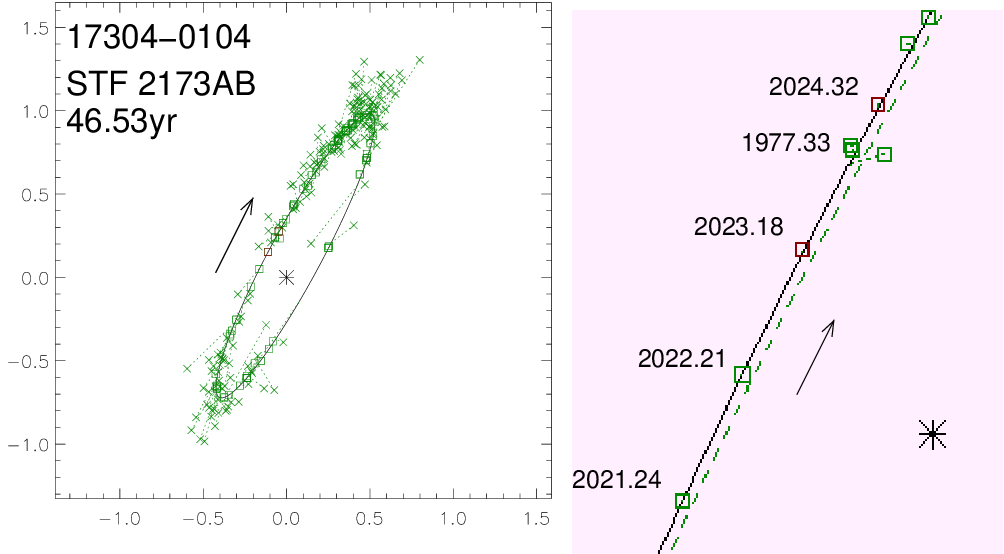}
\caption{The accurate speckle-only  orbit of 17304$-$0104 (STF  2173) on the
  left, with the visual  measurements in the 19th  century overplotted as
  crosses. Its fragment on the right shows the systematic residuals to the
  previous orbit (dashed line).
\label{fig:accurate2} }
\end{figure}

Orbits of classical  visual binaries are computed  and recomputed many
times until  they become  ``definitive''. Even then,  some adjustments
may   be    needed.   Two   relevant   cases    are   illustrated   in
Figures~\ref{fig:accurate1} and \ref{fig:accurate2}.

The  pair 14160$-$0704  (HU  138)  has been  discovered  by Hussey  in
1900.3; by now,  it has almost completed one revolution  of its 143 yr
orbit.  The latest orbit by \citet{Doc1990d} had a period of 151.4 yr.
Accurate positions  are available for the  period 1989.3--2024.15. The
magnitude difference  measured at  SOAR is small,  about 0.2  mag, but
sufficient  to  establish  that  the   secondary  is  located  in  the
south-west quadrant,  differing from  the ephemeris by  180\degr.  The
relative position of the companion in Gaia  DR3 confirms the  quadrant, with
$\Delta G =  0.13$ mag.  So, the Docobo's orbit  should be ``flipped''
(change  $\Omega$ by  180\degr),  and all  visual  measures should  be
flipped   as  well.    Alternatively,   we  can   keep  the   historic
identifications of  the primary  and secondary  stars in  this system,
flip  the modern  measures, and  assign  them a  negative $\Delta  m$.
Apart from  the flip of the  Docobo's orbit, one notes  its systematic
deviation  from  the  latest   measures.   This  justifies  the  small
correction of the elements made  here.  The deviant speckle measure in
1996.42 was given a reduced weight.  The period of 143.0$\pm$2.7 yr is
well established,  but in  the future  this ``definitive''  orbit will
need further minor adjustments as the visual coverage is progressively
replaced  by accurate  relative positions.   Until then,  the historic
visual measures will remain critical for constraining this orbit.

The classical visual binary 17304$-$0104  (STF 2173) was discovered by
W.~Struve in 1829 at 0\farcs6  separation and since then has completed
four revolutions  of its 46  yr orbit.   The latest adjustment  of the
elements  using   all  available   data  was  published   recently  by
\citet{SAM22}.   As shown  in Figure~\ref{fig:accurate2},  right, this
orbit  (dashed  line) has  minor  but  systematic residuals  from  the
speckle  measures, which  started in  1977.3  and now  cover one  full
revolution.  I corrected the orbit using only the 89 speckle positions
and the 29 RVs from  \citet{Pourbaix2000}.  The weighted rms residuals
are 4.4\,mas, the period is  46.538$\pm$0.019 yr.  In this case, using
a large  number of visual  positions, even with low  weights, degrades
the orbit accuracy instead of improving it.  If I include, for a test,
several  early visual  measures in  the LS  fit, the  error of the period
decreases  insignificantly,  to  0.018   yr  (the  full-orbit  speckle
coverage  constrains  the  period  quite well).   The  left  panel  of
Figure~\ref{fig:accurate2} illustrates  the situation  by overplotting
visual measures  made before 1900. Effectively,  these data contribute
only  noise.  So,  when  the micrometer  measures  are not  absolutely
necessary  for the  orbit calculation  (as is  the case  here), it  is
better to ignore them.

\subsection{Speckle-Only Orbits}
\label{sec:speckle}

\begin{figure*}[ht]
\epsscale{1.0}
\plotone{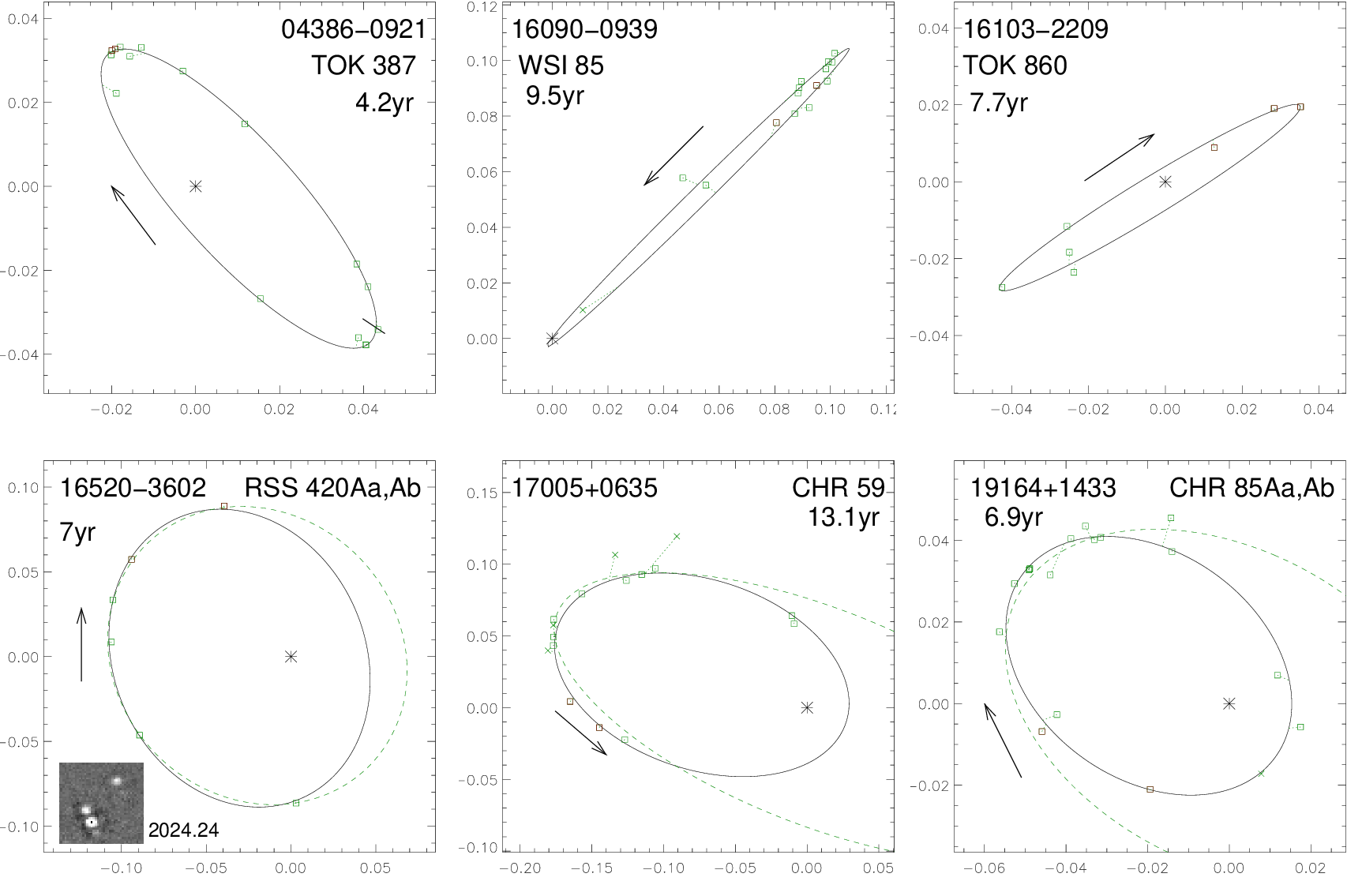}
\caption{Six orbits based only on speckle measurements. The WDS codes,
  discoverer designations, and periods are indicated. Alternative orbits
  are plotted by dotted lines.
\label{fig:speckle} }
\end{figure*}

On average, binary stars discovered by speckle interferometry are
closer than classical visual binaries. So, they move faster, allowing
calculation of orbits with periods on the order of a decade. Six such
orbits are illustrated in Figure~\ref{fig:speckle} and commented
below. 

{\it 04386$-$0921.}  The pair  was resolved  at SOAR  in 2014;  it has
completed 2.4 revolutions,  and the orbit with $P=4.239  \pm 0.035$ yr
determined  from  16 measures  made  exclusively  at SOAR  is  tightly
constrained. The  rms residuals  are only 1  mas. The  small magnitude
difference $\Delta  m \sim 0.5 $  mag and the spectral  type F8V favor
measurements  of   the  RVs  for   both  components,  so   a  combined
spectro-interferometric orbit can  be determined in the  future if the
pair is monitored spectroscopically.

{\it 16090$-$0939.}  This bright ($V=7.24$ mag) A8V star HIP~79122 was
resolved for the first time in 2008  at the 4 m Blanco telescope using
HRCam \citep{TMH10}.  A substantial  magnitude difference of $\Delta m
\sim  3.5  $  mag  and  the large  eccentricity  conspire  to  prevent
resolution of  this pair near the  periastron (the cross in  the orbit
plot marks  the non-resolution in  2021.3), so this important  part of
the orbit is not covered and the eccentricity is not well constrained.
In the orbit fit, I fixed $e=0.97$ to get a mass sum of 5.3 \msun with
the  Gaia  DR3  \citep{EDR3}   parallax  of  12.57$\pm$0.09  mas.   An
unconstrained  fit  converges  to  $e=0.99$  and  corresponds  to  the
excessive mass sum  of 16 \msun.  The residuals are  2.2 mas.  A large
RV variation around the next periastron in 2030.4 is expected.

{\it  16103$-$2209.}   This star,  HIP~79244  (spectral  type A0V),  was
resolved at SOAR in 2019.  Six measurements made to date (plus two
non-resolutions in 2022)  constrain the 7.7 yr  orbit reasonably well;
it  yields a  mass sum  of 7.6  \msun with  the Gaia  DR3 parallax  of
6.26$\pm$0.09 mas.  In 2023.18, the  separation of 15.5 mas  (under the
diffraction limit) was estimated by fixing $\Delta m$. Additional
measurements are needed to confirm and improve this orbit.  

{\it 16520$-$3602.}   This relatively  faint ($V=12.0$ mag)  M2V dwarf
HIP~82521 (NLTT~43673) was  resolved in 2019 into a  tight triple with
separations    of   0\farcs09    and   0\farcs4    (the   insert    in
Figure~\ref{fig:speckle} shows  the latest shift-and-add image  of the
triple).   In the  WDS,  these  pairs received  the  DDs 
designations RSS~420Aa,Ab and RSS~420AC, respectively.  Note, however,
that the faint companion B at 6\farcs6 which gave rise to the original
RSS designation  is unrelated according  to the Gaia  astrometry.  The
separations in this triple and the Gaia DR3 parallax of 26.76$\pm$0.31
mas  imply the  inner and  outer orbital  periods $P^*$  of a  few and
$\sim$50 yr, respectively.  After five  years of monitoring, the first
7 yr  orbit of the  inner pair  can be determined.   The unconstrained
orbit fit gives $e=0.22 \pm 0.11$ and a mass sum of 0.64 \msun (dashed
line  in  Figure~\ref{fig:speckle}),  while  the  luminosity  and  the
spectral type  of these stars correspond  to a mass sum  of 1.0 \msun.
By enforcing $e=0.4$, the mass sum  is brought into agreement with the
expectation, while the  residuals do not increase.  In  the coming few
years, the inner  orbit of this triple will  become fully constrained,
but monitoring  of the outer pair  A,C must continue.  Both  inner and
outer  pairs in  this low-mass  triple system  have retrograde  motion
suggestive of mutually aligned orbits.
 
{\it   17005+0635.}   This   bright  A7V   star  HIP~83223,   resolved
interferometrically     by    the     CHARA     team    in     1985.52
\citep{McAlister1987a},  is designated  as CHR~59.   The latest  orbit
with $P=26.49$ yr was published by \citet{SAM21}.  It assumed that the
CHARA observations  in the 1990s  and the  modern SOAR data  cover two
opposite  ends   of  a   near-circular  orbit.   However,   the  CHARA
observations  did not  constrain  the quadrant  (it  can be  flipped),
despite  the substantial  magnitude  difference of  2.3  mag.  The  AO
measure by  \citet{Roberts2018} in  2004.5, while  inaccurate, clearly
indicates that the companion was resolved always in the same quadrant,
hence the orbital period is two times shorter than assumed previously.
This drastically revised  orbit with $P = 13.11 \pm  0.06$ yr fits the
data better than  the previous 26 yr orbit.  It  corresponds to a mass
sum of  3.0 \msun with  the Hipparcos parallax of  13.26$\pm$0.47 mas.
The Gaia  DR3 parallax of  15.06$\pm$0.21 mas, although  formally more
accurate, is likely biased by the  fast motion near the periastron (in
2014.44) and gives an even smaller mass sum of 2.0 \msun. Observations
near  and after  the  next periastron  in 2027  will  help to  further
constrain this orbit.

{\it 19164+1433.}   This early-type (HD  185055, B9V, $V =  5.63$ mag)
pair was  also resolved  in 1985  by CHARA in  their survey  of bright
stars  \citep{McAlister1987a}.   The  orbit   with  $P=13.67$  yr  was
published by \citet{TMH15}. Here it  is revised drastically to $P=6.9$
yr. The  quadrants of the  SOAR measures  are known, showing  that the
pair  is  always resolved  in  the  same  quadrant,  and the  orbit  is
eccentric.   The  situation  resembles  the previous  case,  but  here
observations near  the periastron  are actually  available, confirming
the  eccentric  orbit.  Gaia  DR3  measured  an accurate  parallax  of
6.244$\pm$0.034 mas for the tertiary  companion B located at 8\farcs3,
while the astrometry of the inner binary is biased by the fast orbital
motion. With this parallax, the new  orbit corresponds to the mass sum
of 5.2 \msun.

\subsection{Spurious Visual Pairs: HD 129732}
\label{sec:fake11}

\begin{figure}[ht]
\epsscale{1.1}
\plotone{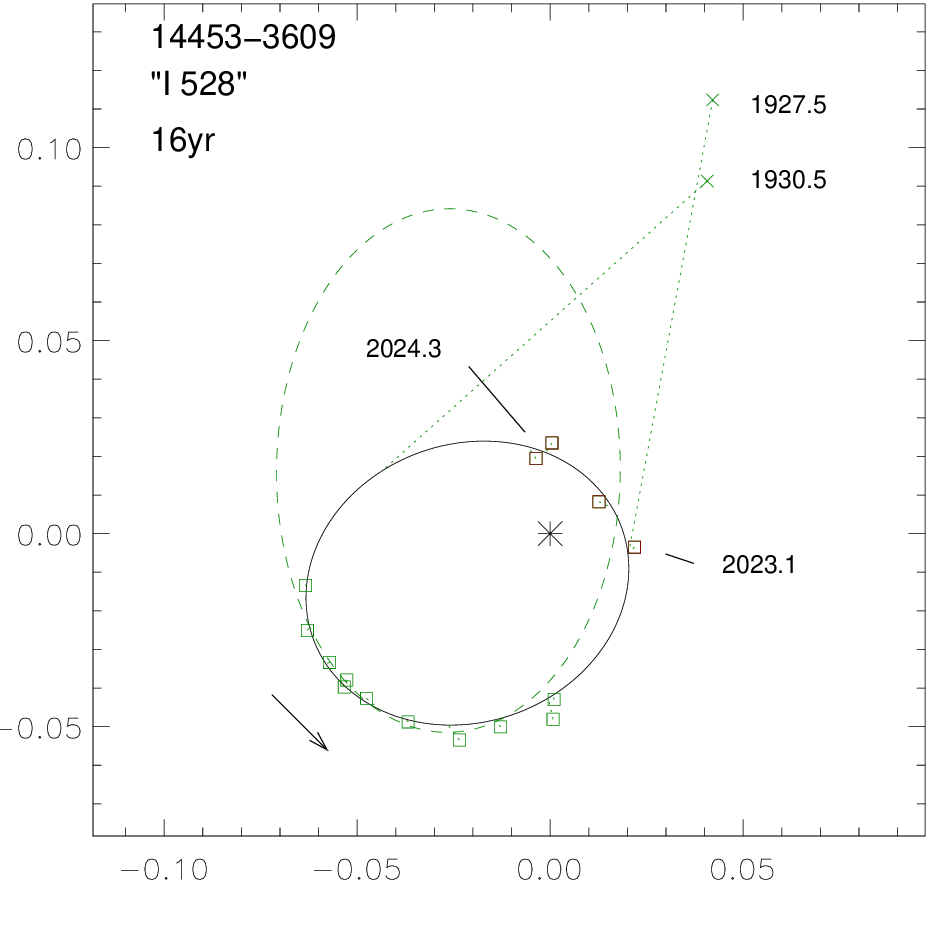}
\caption{Orbits of 14453$-$3609 with periods  of 15.9 yr (full line) and
  31.8 yr  (dash). Two discrepant  visual measures are  plotted by
  crosses and connected to the ephemeris positions by dotted lines.
\label{fig:sp1} }
\end{figure}

The star  HD 129732  (A1IV, $V=7.37$ mag)  was presumably  resolved at
0\farcs4 by R.~Innes in 1909 and  it was designated as I~528.  The WDS
database contains two more visual resolutions in 1927 and 1930 at much
closer separations of 0\farcs1 and several visual non-resolutions.  In
the modern  epoch, the pair  remained unresolved by Hipparcos  and was
twice  (in 1989  and  1993)  resolved by  the  CHARA  team. The  HRCam
measures  cover  the  2008--2024  interval (in  2008.5  the  pair  was
unresolved).  An  orbit with $P=33$ yr  was fitted to the  measures in
2018 and  slightly refined by  \citet{SAM21}.  In the light  of recent
observations  near the  periastron (Figure~\ref{fig:sp1}),  this orbit
appears  to be  incorrect,  and the  actual  period is  15.93$\pm$0.16
yr. All three  visual measures turn out to be  spurious; in fact, this
pair  {\em never} has been resolved visually.  An attempt to match the two
misleading visual  measures is  one of the  reasons for  computing the
previous (wrong) orbit.

Speckle observations reveal many cases of spurious visual resolutions,
sometimes ``confirmed'' by several  subsequent measures. Lists of such
spurious  pairs are  published regularly  \citep[e.g.][]{SAM22}. Here,
the star in question is indeed  binary, but its visual resolutions are
nevertheless   spurious.   Another   example   is  HIP~104440   
(HD~200525, I~379, WDS 210094$-$7310),  presumably discovered by Innes
at 1\arcsec.  It has  an eccentric  visual-spectroscopic orbit  with a
maximum separation  of 0\farcs3  \citep{CHI23}.  The ORB6  catalog even
contains  two {\em  orbits of  single stars}  based on  their multiple
spurious  visual  measures!   Both  pairs,  104~Tau  (WDS  05074+1839,
A~3010)  \citep{Tok2012a}  and  HD~21161  (WDS  03244$-$1539,  A~2909)
\citep{Tok2019},  were ``discovered''  by R.~Aitken,  one of  the most
experienced and trusted visual observer of the past.

\subsection{Spurious Visual Pairs: HR 6100}
\label{sec:fake2}

\begin{figure}[ht]
\epsscale{1.1}
\plotone{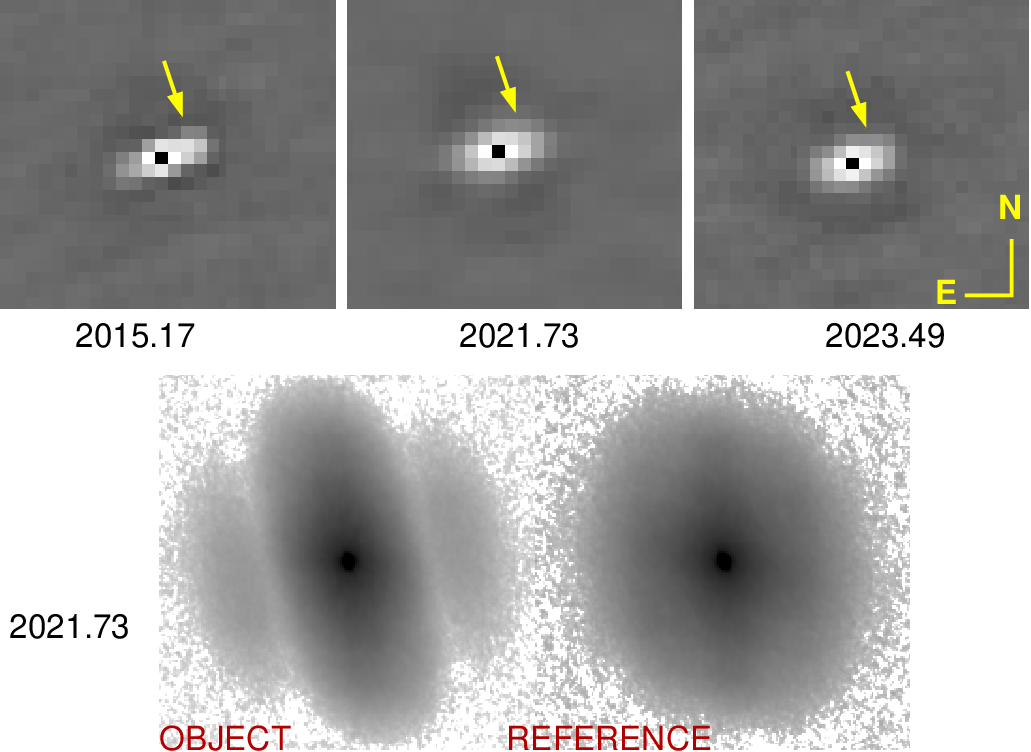}
\caption{Top  row: three  shift-and-add  imagesof  HR 6100  suggesting
  resolutions in the  same quadrant (black dots mark  the center). The
  speckle power spectra of the object  and the reference star taken in
  2021.73 are shown below.
\label{fig:SAA} }
\end{figure}

The bright  ($V=5.42$ mag, B8V)  star HR  6100 (HD 147628,  HIP 80390)
belongs to  the Upper Scorpius  association with  an age of  8--20 Myr.
The  Gaia DR3  parallax  of  7.65$\pm$0.12 mas  is  close  to the  DR2
parallax   of   7.89\,mas   and   to   the   Hipparcos   parallax   of
7.75$\pm$0.25\,mas.  So,  the binary  nature of the  source apparently
did not spoil its astrometry.

In 1927,  \citet{vdB1928} resolved this star  as a visual binary  at a
separation of  0\farcs11, with nearly  equal components.  It  has been
measured in the period 1927--1947 three  times by himself and twice by
W.~Finsen, apparently confirming this  discovery.  The pair is denoted
as B~868 and  16245$-$3734.  However, the WDS  database also documents
13  non-resolutions  (or  partial  resolutions)  by  visual  observers
between 1936 and 1968.  The  early speckle interferometry resolved the
binary, with three measures in 1989 and 1991 at separations between 30
and 56 mas  \citep{McAlister1990,Hartkopf1993,Hartkopf1996}.  The pair
was not  resolved by Hipparcos, hence  around 1991 it was  closer than
0\farcs1.

\begin{figure}[ht]
\epsscale{1.1}
\plotone{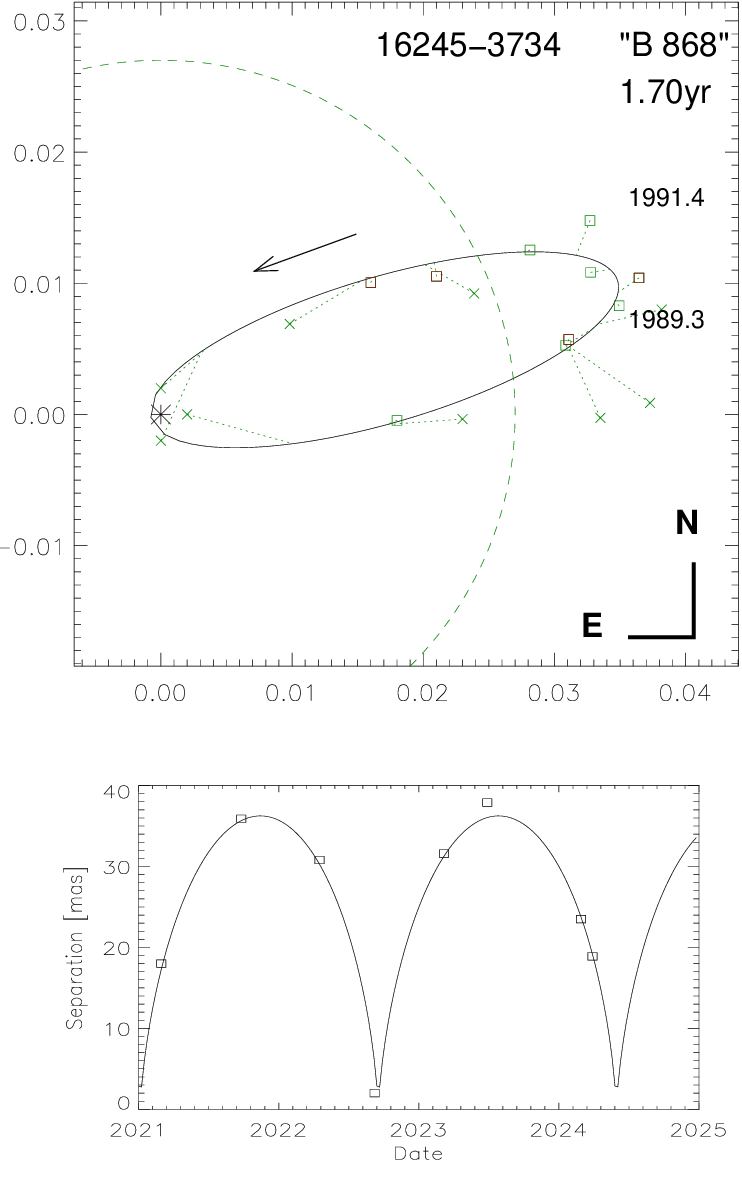}
\caption{the orbit of HR~6100. Accurate  positions are plotted as squares,
  less   accurate   as   crosses  (including   three   non-resolutions
  arbitrarily  placed  at  2\,mas   separation).   The  dashed  circle
  indicates the  diffraction limit of  27\,mas.  The lower  plot shows
  separation vs. time.
\label{fig:B868} }
\end{figure}

Starting  from  2008, this  bright  and  close binary  was  frequently
visited by  HRCam using the  $y$ filter (central  wavelength 540\,nm).
The magnitude difference of $\sim$0.5 mag allows identification of the
correct  quadrant from  the  shift-and-add (SAA)  images  at times  of
maximum  separation.    Figure~\ref{fig:SAA}  gives  the   three  best
examples  indicating  that  in   2015.17,  2021.73,  and  2023.49  the
companion was located  on the upper-right side, in  the same quadrant.
In 2015  the data  were recorded  with a  5 ms  exposure time  and the
speckle contrast was higher than with  the 25 ms exposures used later.
At  maximum separation,  the second  fringe in  the power  spectrum is
clearly detected, allowing very  accurate position measurements (using
a reference  star), but  at separations  below the  formal diffraction
limit of  27\,mas, only the  central fringe is  present.  Measurements
under the  diffraction limit are  less accurate, they are  obtained by
fixing  $\Delta m$.   Larger errors  are  also assigned  to the  HRCam
measurements made without reference stars.   In three visits, the pair
was securely unresolved, indicating separations well below 10\,mas.

The HRCam data  allow calculation of an eccentric orbit  with a period
of 1.70 yr, shown in Figure~\ref{fig:B868}. According   to this orbit,
the separation never exceeds 38\,mas, just above the diffraction limit
of 4 m telescopes.  Three secure non-resolutions at SOAR correspond to
the epochs near periastron.  If  some position angles are flipped, the
data can be fitted by an orbit with a small eccentricity and two times
longer period,  but such orbit  contradicts the known  quadrants.  The
free fit converges  to a face-on orbit with $i=0$,  so the inclination
is fixed to a small but  more probable value of 20\degr.  The weighted
residuals of  1.6\,mas match the  estimated measurement errors,  but a
substantial uncertainly in the elements $a, \Omega, \omega, i$ remains
because the periastron is not covered.   The orbit gives a mass sum of
5.5 \msun (parallax 7.65\,mas), close to the masses estimated from the
absolute magnitudes.  However, adopting a larger eccentricity leads to
a  larger  mass  sum,  so  the  orbit  does  not  yet  provide  useful
constraints on the mass.  Rather,  it does not contradict the expected
masses and distance. Accurate masses can be measured only by observing
this pair near the periastron with long-baseline interferometers.

\begin{figure}[ht]
\epsscale{1.1}
\plotone{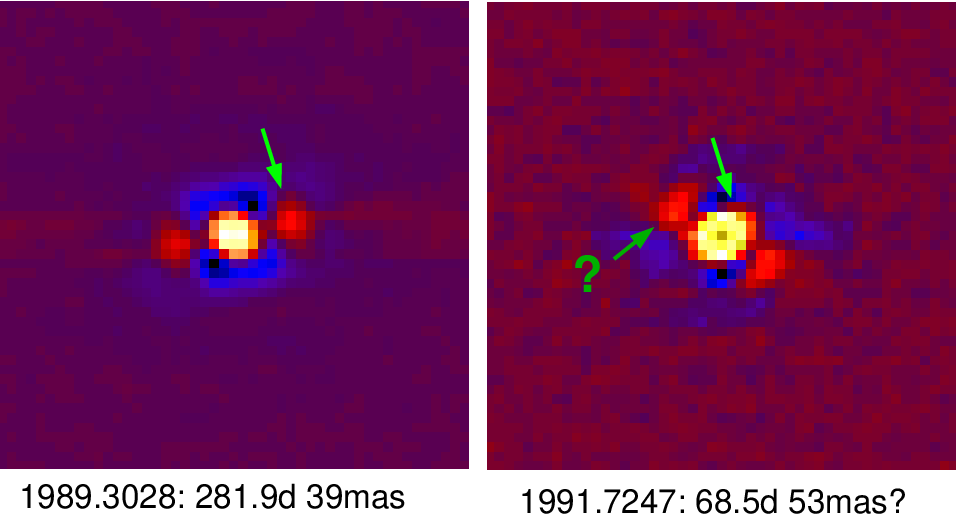}
\caption{Speckle  DVA   images  of HR 6100 from   the  CHARA   archive  (courtesy
  B.~Mason). The green arrows mark the companion's location according to the
  orbit.   The darker  green arrow  with a  question mark  indicates the
  published position in 1991.72.
\label{fig:CHARA} }
\end{figure}

Of the  three speckle measures  published by  CHARA, the first  one in
1989.30 fits the orbit, the measure  in 1991.39 is marked as uncertain
and can be  ignored, and the measure in  1991.72 (68\fdg5, 0\farcs053)
is in strong disagreement; the orbit predicts 295\fdg0, 0\farcs029. On
my  request,  B.~Mason  reviewed these  data.   Figure~\ref{fig:CHARA}
shows  the Directed  Vector Autocorrelation  (DVA) images  recorded by
CHARA at two  epochs.  In both, the companion is  separated by $\sim$6
image pixels, but the published separations  of 39 and 53 mas indicate
a  coarser pixel  scale in  1991.72.  On  the 1991.72  DVA image,  the
central and side peaks are elongated in the direction predicted by the
orbit.  Apparently,  the pair, separated  by 29\,mas according  to the
orbit, has  not been resolved, while  the measured peak at  59\,mas is
spurious.  Similar  spurious details, called optical  ghosts, are also
documented in the HRCam data  \citep[see Figure 11 in][]{TMH10}.  They
are  distinguished from  the  real companions  by observing  reference
stars and by comparison with observations of other stars made close in
time.
  
The reported visual resolutions of HR 6100 appear to be spurious.  With
the  0.7  m  telescopes  used  by  the  visual  observers,  the  image
elongation is always totally negligible. As noted above, this is not a unique
case.  Hypothetically, the eccentric orbit  of this young pair could shrink
through energy dissipation near periastron  via tides or friction with
the residual  gas.   If the  semimajor  axis  has shortened  substantially
during  $\sim$100  yr,  the  visual resolutions  could  be  explained.
However, such a  fast orbit decay implies the  energy dissipation rate
exceeding the luminosity  of the stars by a large  factor of $\sim$30,
while in fact the luminosity is  normal and matches the age, and the
brightness  is  constant.   \citet{Sharma2022}  discovered  only  minor
photometric pulsations  of this  object with  frequencies of  2.54 and
2.59 cycle~day$^{-1}$, typical  of rapidly rotating ($V \sin  i = 160$
\kms) B-type stars.

\subsection{Drastic Orbit Revisions}
\label{sec:drastic}

\begin{figure*}[ht]
\epsscale{1.0}
\plotone{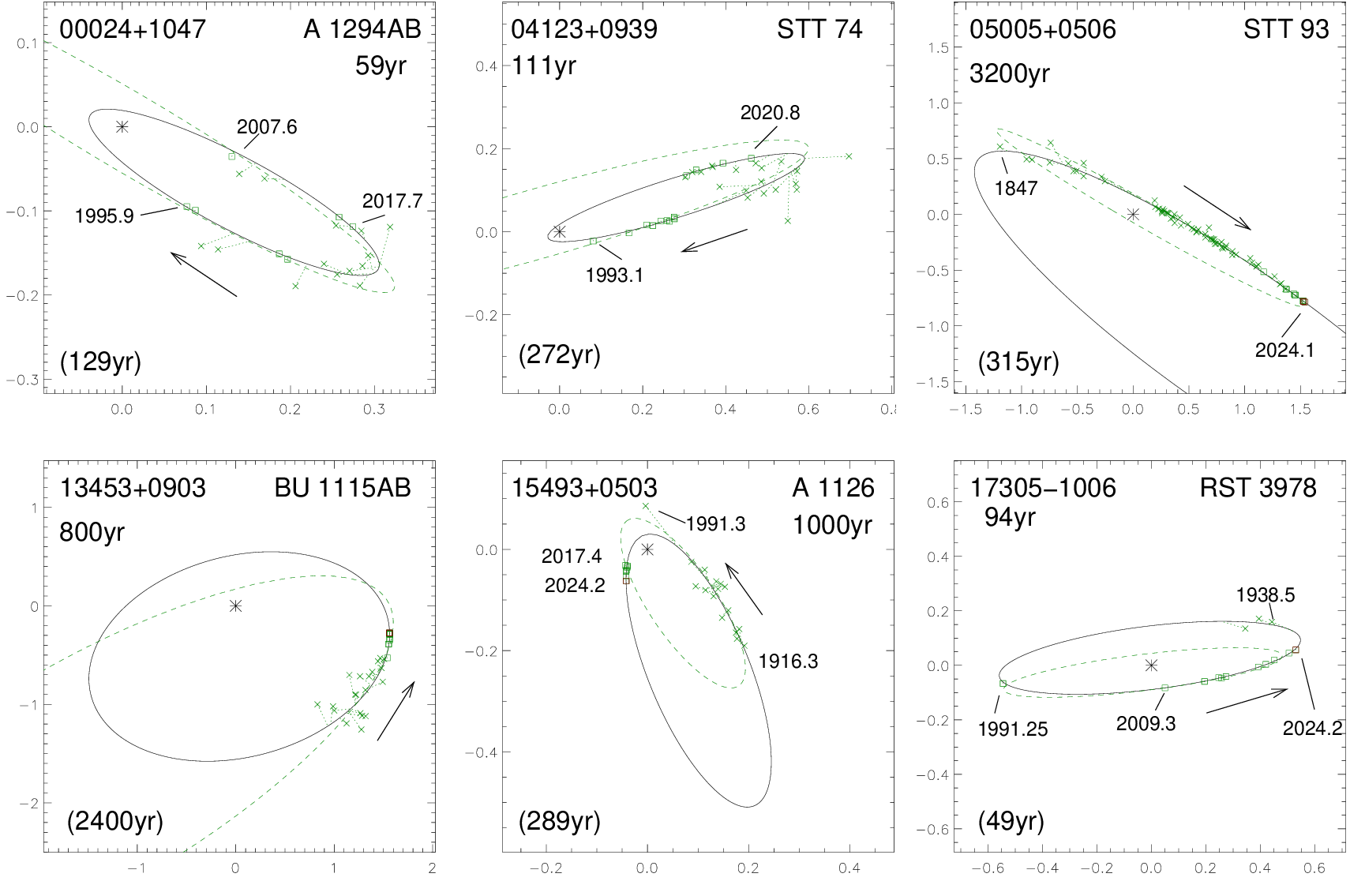}
\caption{Six  strongly  revised  orbits.  All these pairs  were  discovered
  visually.   Their WDS  codes, discoverer  designations, and  revised
  periods  are indicated,  while  periods of  the previously  computed
  orbits are given in brackets.
\label{fig:drastic} }
\end{figure*}

In this Subsection, examples of  substantial orbit revisions are given
(Figure~\ref{fig:drastic}).  All  orbits are based on  the combination
of  visual and  speckle  measurements and  have  periods ranging  from
decades to centuries. The reasons of the revision are detailed below.

{\it  00024+1047.}  The  pair A~1249AB  (HIP 190)  passed through  the
periastron in 2003 and is  presently approaching the apastron.  Almost
two 59 yr orbital cycles are  covered since its discovery by Aitken in
1905.   The previous  129 yr  quasi-circular orbit  by \citet{Zir2003}
flipped the  speckle measures made  by CHARA and all  visual measures,
although with  the magnitude difference  of $\Delta  m = 0.8$  mag the
flips of the visual measures are questionable.  The tertiary companion
C  at 63\arcsec  ~(HIP  185)  has an  accurate  Gaia  DR3 parallax  of
10.08$\pm$0.15\,mas.  The mass sum of  A,B according to the new orbit,
2.26 \msun, matches the spectral type G0 and the masses estimated from
the luminosity better  than the Zirm's orbit, which yields  a mass sum
of 2.8 \msun.

{\it 04123+0939.} O.~Struve resolved the  F0 star HD~26547 at 0\farcs4
in  1849 \citep{Struve1878}.  The pair  has closed  down and  remained
unresolved  until  1894;  its  slow widening  and  closing  at  nearly
constant angle is  well documented by visual observations  in the 20th
century.  The  magnitude difference of $\Delta  m = 1.2$ mag  does not
allow  quadrant  flips.   The  CHARA speckle  observations  cover  the
approach to  the periastron in 1996,  and presently the pair  opens up
again.   So, the  previous  272  yr orbit  by  \citet{Msn2019} is  not
correct.  No measurements of the parallax are available.


{\it 05005+0506.} This G2/3V star  HIP 23277 was resolved by O.~Struve
in 1847 at 0\farcs82. It moves  slowly on a slightly curved segment of
its    long-period   orbit    and    presently    is   at    1\farcs73
separation. Although the observed arc  covers nearly 180\degr, it does
not constrain  the orbit, so I  fixed the period and  the eccentricity
and fitted the  remaining elements. Gaia DR3 measured both  stars at a
separation of  1\farcs278, position angle 246\fdg2,  $\Delta G= 0.606$
mag (in  2016.0) with  matching parallaxes of  14.94$\pm$0.03\,mas and
15.02$\pm$0.07\,mas.  The  parallaxes and the new  orbit correspond to
the mass  sum of 2.0 \msun,  in agreement with the  spectral type.  In
contrast,  the   315  yr   orbit  computed  by   \citet{Izm2019}  from
essentially the same data yields  an unrealistically large mass sum of
16.7\msun  and   deviates  from  the  latest   accurate  measurements.
Although the new orbit is poorly constrained, it accurately models the
observations  and   can  be   used  for  retroactive   or  near-term
calibration of  speckle measures.

{\it  13453+0903.}   This pair,  HIP~67115  (G5,  parallax 16.70  mas)
resembles the previous  one: only a short arc of  its orbit is covered
by the existing measures from 1873 to 2024. Here, the previous
orbit computed  by \citet{Izm2019} is  revised in the  opposite sense,
from the long 2400 yr to a  shorter 800 yr period. The reason for this
revision  is the  mass sum,  reduced from  8.8 to  1.9 \msun,  and the
systematic residuals of the accurate measures to the old orbit. Again,
I fixed $P$ and $e$ and  fitted the remaining elements.  This orbit is
also a  good calibrator of pixel  scale and orientation tied  to Gaia,
which measured both components.

{\it 15493+0503.} The pair  HIP~77489 (K0, parallax 2.60$\pm$0.04 mas)
has a visual  coverage from  1905 to  1966, and  during this  time the
separation slowly decreased.  One speckle  measure by CHARA in 1991.32
is available, and the segment  with increasing separation between 2017
to  2024 (after  passage through  the periastron)  is well  covered by
seven HRCam  measures.  The  small $\Delta  m$ allows  quadrant flips.
The 289 yr orbit by  \citet{Gomez2022} substantially deviates from the
recent  measures, calling  for its  revision (see  the dashed  line in
Figure~\ref{fig:drastic}).  I  could not reconcile the  1991.3 speckle
measure  with the  latest observations  and prefer  to ignore  it. The
orbit is poorly constrained, so I  fixed its period. The mass sums for
the  old and  new orbits  are 5.2  and 1.4  \msun, respectively.   The
latter value matches  the spectral type and the  absolute magnitude of
$M_V \approx 5.8$ mag, suggesting that the stars are not evolved.

{\it 17305$-$1006.}  HIP~85565 (G8V,  parallax 19.96$\pm$0.44 mas) was
resolved   by   R.    Rossiter   in  1938   at   110\fdg2,   0\farcs47
\citep{Rst1955}.  Only  three visual  measures were made  by Rossiter,
showing little  motion till 1951.   The magnitude difference of  2 mag
should not allow quadrant flips. The Hipparcos and HRCam data together
cover  almost half  of  the  orbit and  indicate  a  period around  90
yr. Yet, the  previously computed orbit \citep{TMH15} had  a period of
49 yr, postulating that the visual  and Hipparcos positions are in the
same quadrant  and that the  pair has made one  revolution in-between.
This orbit, however, yields an implausible mass sum of 11.7 \msun.  In
the revision,  I flipped the  three visual measures instead  of simply
ignoring them and obtained a mass sum of 2.7 \msun.  The quadrant flip
of the visual  measures can possibly be justified by  the human nature
of the  observer.  If the  first measure  was attributed to  the wrong
quadrant by  error, subsequent measures  made by the same  person were
placed in the same quadrant in order to be consistent.

\subsection{Tentative Orbits}
\label{sec:tent}

\begin{figure*}[ht]
\epsscale{1.0}
\plotone{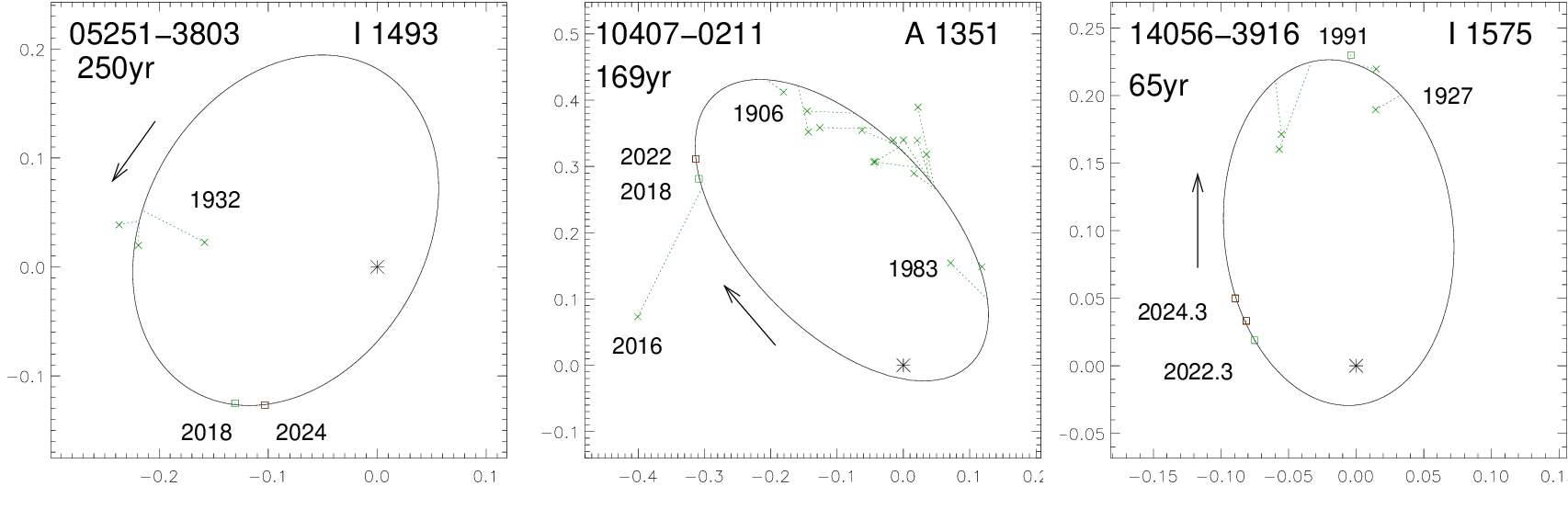}
\caption{Three poorly constrained orbits.
\label{fig:tentative} }
\end{figure*}

The previous Section contained some  poorly constrained long-period orbits
based  on short  arcs. Here,  another flavor  of first-time  tentative
orbits  based  on  scarce  data   is  illustrated  by  three  examples
(Figure~\ref{fig:tentative}).

{\it 05251$-$3803.}  The  250 yr orbit of HD~35724 based  on only five
positions  (three  visual  and  two  speckle)  represents  an  extreme
case. The discovery measure by Innes in 1927 is ignored because of the
largely discrepant separation, and the  remaining three visual measures in
1932--1937 do not show any significant motion.  The two HRCam measures
in 2018  and 2024 indicate that  the motion is direct.   These clearly
insufficient data match an orbit with  $P=250$ yr.  By fixing also the
eccentricity $e=0.7$, I  obtain a converging LS fit and  a mass sum of
3.0  \msun  (parallax 4.17$\pm$0.32  mas,  spectral  type F3V).   This
``guesstimate''  orbit predicts  that by  2031 the  pair will  turn by
10\degr  ~and will  close down  to 0\farcs12,  accelerating toward  the
periastron in 2045.

{\it 10407$-$0211.}   The visual  pair HD~92484  (G5) was  resolved in
1906 at 0\farcs45  by R.~Aitken.  A preliminary  orbit with $P=169$~yr
is proposed  here.  The visual measures  until 1983, when the  pair became
too  close,  show a retrograde  motion.  The  two HRCam  measures  after
passage  through the periastron in  1991  are  attributed to  the  second
quadrant  (with  $\Delta   m  \approx  0.15$  mag,   the  quadrant  is
uncertain).  An alternative orbit with a longer  period is possible if the
HRCam measures  are flipped. As  the orbit segment near the periastron is
not covered, I fixed $e=0.9$ in the LS fit.

Gaia DR3 contains two sources at 0\farcs4073 separation with $\Delta G
=  -0.006$  mag; neither  of  them  has 5-parameter  astrometry.   The
separation is close to the  orbit prediction, while the position angle
is clearly discrepant.   The orbit predicts 48\fdg58  in 2016.0, while
Gaia  measured  79\fdg66.   A  likely  reason of  the  Gaia  error  is
misidentification  between the  two sources  in different  scans.  The
resulting  discrepant abscissae  do not  yield acceptable  5-parameter
astrometric models.   Indirectly, this suggestion is  supported by the
near-zero  magnitude difference:  if  the sources  are swapped  frequently,
their  mean  fluxes will  be  nearly  the  same.  Source  swapping  is
mentioned by \citet{Holl2023}, who discuss  various caveats of the Gaia
data related to non-point sources.  The comparison of the HRCam calibrator
binaries  with  Gaia  DR3 \citep{SAM21}  revealed  several  discrepant
pairs;  in all  of  those,  one or  both  components  lacked the  full
5-parameter astrometric solutions in  DR3, indicating problematic Gaia
astrometry.

{\it  14056$-$3916.} HIP~68831  (G3V,  parallax  6.82$\pm$0.15 mas)  was
resolved for  the first time at  0\farcs22 in 1927 by  Innes. The pair
was apparently  unresolved visually after 1934,  re-appearing again in
1960 (I  ignored this discrepant  measure) and measured twice  in 1979
and 1985.   The quadrant  of the Hipparcos  measure agrees  with these
visual resolutions; no  measures were made by CHARA.   The three HRCam
points show  increasing separation  after the periastron passage  in 2018.
Overall,  the orbit  looks as  almost  constrained (I  fixed only  the
inclination) and leads to the mass sum of 1.9 \msun.


\section{Discussion, Recommendations, and Trends}
\label{sec:rec}

In  this  work, the  challenge  of  the  visual orbit  calculation  is
illustrated  by examples  ranging  from trivial  minor adjustments  to
radical  revisions and  tentative  poorly constrained  orbits. In  all
cases, understanding the  nature and reliability of the  input data is
crucial. The  system of  relative weights and  grades adopted  in ORB6
\citep{VB6} is  rooted in  the epoch of  visual measures  and nowadays
becomes  obsolete.   Excessive  weight  ascribed to  the  visual  data
(compared  to speckle)  degraded the  quality of  orbits published  in
\citet{TMH15}, so  many of  these orbits  had to  be revised  later to
bring them into agreement with  the accurate modern measures (two such
examples are found  in Table~1).  A large effort  of orbit calculation
undertaken by  \citet{Izm2019} was  also based  on the  formal fitting
without prior manual data inspection and without a sanity check of the
mass sum; consequently, some of  those orbits also require substantial
revisions (Figure~\ref{fig:drastic}).

The database  of historic  visual double star  measures is  an amazing
heritage,  allowing  us  to  determine  orbits  with  periods
measured in centuries.   However, the existing visual  data cover only
relatively  bright pairs.   Whenever  their quality  and quantity  was
adequate, the orbits were already computed, and only their improvement
is  possible by  continued speckle  monitoring.  The  remaining visual
pairs without  orbits typically  have insufficient historic  data, and
even when  they are complemented  by several modern  observations, the
resulting   orbits  are   poorly  constrained   (e.g.  14407$-$0211   in
Figure~\ref{fig:tentative}).   So,  the  treasure  trove  of  the  old
observations is not very deep and eventually it will be exhausted.  We
are indebted  to R.~Gould who  explored the  WDS content in  search of
pairs where modern  observations could lead to an  orbit.  This effort
in double  star archaeology, in  combination with the HRCam  data, has
led to many new orbits, although  some of those are tentative owing to
the scarce data.

My  experience in  computing orbits  based on  visual  and speckle
measures is summarized in the following recommendations.
\begin{itemize}
\item
When  the accurate  (speckle or  other)  data are  available, the  old
visual measures should  not be used.  Effectively, they  add noise and
degrade the quality of the orbit, instead of improving it.

\item
Historic  visual  measures should  be   used with  caution.  The
separations are often strongly biased. The number of wrong resolutions
(when  the pair  was in  fact  too close  or  the star  is single)  is
substantial, as  well as  the number  of discrepant  measures.  Manual
inspection  and editing  (selective  rejection  or down-weighting)  of
visual data  is mandatory in all  cases. When both visual  and speckle
measures  are used,  they must  be  weighted according  to their  real
errors.

\item
Not   all    speckle   data   are   correct,    their   rejection   is
acceptable. There are some spurious speckle resolutions.

\item
The accuracy of the speckle data depends  not only on the aperture size of
the  telescope,  but  also  on the  magnitude  difference,  magnitude,
used equipment, measurement method, and the team that produced the data.

\item
For pairs  closer than $\sim$2\arcsec, relative  positions provided by
Gaia can  be incorrect,  possibly owing to  confusion between  the two
sources. Existence  of the 5-parameter astrometric  solutions for both
components of a pair is  necessary for trusting the relative positions
in DR3.


\end{itemize}

The classical epoch, when an implicit  goal was to compute an orbit of
every pair with observed motion, is  coming to an end. Below I outline
the future trends in the orbit calculation.

The Hipparcos  mission discovered thousands  of close pairs  missed by
the visual surveys. A rapid  orbital motion was expected and confirmed
for  close  Hipparcos binaries  in  the  solar neighborhood,  and  the
follow-up      speckle      effort     produced      their      orbits
\citep[e.g.][]{Horch2015}.   Similarly,  fast  movers  among  recently
resolved  nearby  stars  (mostly   low-mass  dwarfs),  when  regularly
monitored,    allow     determination    of     good-quality    orbits
\citep{Mann2019,Vrijmoet2022}. In all  these cases, we do  not have to
deal with  the historic visual data,  but a good understanding  of the
nature of Hipparcos and speckle measures is still needed for computing
orbits.

Extrapolating  the  Hipparcos  results  to  Gaia  opens  new  exciting
horizons.   The number  of  astrometric orbits  in DR3,  $\sim$10$^5$,
already      largely     exceeds      the     content      of     ORB6
\citep{Halbwachs2023}. However,  those orbits have only  short periods
(less than $\sim$3  years), and a substantial  fraction of low-quality
solutions.  The median  period of solar-type binaries is  of the order
of  $10^5$ days  or  300\,yr \citep{Raghavan2010},  while the  planned
duration of the Gaia  mission is only 10 yr.  So,  even the final Gaia
data  release (DR5)  will  determine astrometric  orbits  for a  minor
fraction of  nearby binaries,  and the orbit  catalogs like  ORB6 will
remain relevant for slower pairs.  Furthermore, a direct resolution of
astrometric binaries is  needed to derive masses even  when their Gaia
orbits are  known   (the mass  sum depends  on the  full semimajor
  axis,  while Gaia  measures only  the axis  of the  photocenter).  
Another  essential ingredient  are  accurate  parallaxes.  For  visual
binaries, Gaia either  does not provide parallaxes  or gives distorted
values  because the  standard  5-parameter  astrometric solutions  are
biased  by  the  orbital  motion \citep{Chulkov2022}.   This  bias  is
removed for  binaries with astrometric orbits  and partially corrected
for  acceleration solutions.   However, when  the individual  transits
become public in DR4 (expected in 2026), their joint analysis together
with relative positions and visual  orbits will de-bias the parallaxes
and lead  to accurate masses.   So, the continuation and  extension of
the ground-based  speckle interferometric monitoring appears  to be an
essential complement to Gaia.

The Gaia mission  will eventually publish millions of  new faint pairs
with separations  above $\sim$0\farcs1 in  the final DR5; at  the same
time,  the  quality  and  quantity of  Gaia  astrometric  orbits  will
improve, compared  to the current  DR3.  The overwhelming  majority of
new Gaia  binaries will be distant  and will move too  slowly. So, the
future effort of  monitoring and orbit calculation  must be selective,
as was the case with the  Hipparcos pairs. The estimated periods $P^*$
and  the periods  of astrometric  orbits  will serve  for selecting  a
subset  of fast  movers among  Gaia  binaries.  Instead  of aiming  to
compute orbits of  all pairs with detectable  motion, future follow-up
observations using  speckle interferometry  and other methods  will be
motivated by the use of these orbits.  For the measurements of masses,
for example,  only a subset of  fast movers will be  sufficient,
while  the requisite  high  accuracy of  position measurements  favors
long-baseline  interferometers  rather  than  speckle  interferometry,
especially for massive and bright  stars.  The new pairs discovered by
Hipparcos,  Gaia, and  high-resolution imaging  surveys lack  historic
measures, so  the time coverage  will be a critical  consideration for
calculation  of  their  orbits.   Orbits of  long-period  binaries  of
interest (e.g.  exohosts or PMS stars) will be estimated approximately
from short observed arcs.

A shift of the paradigm  from holistic to selective observations means
that the orbital motion of the majority of binaries (both historic and
new) will  no longer be monitored.   If, in the future,  some of those
systems become interesting, they will lack the time coverage for orbit
calculation.   This is  particularly sensitive  for orbits  with large
eccentricity.   If   we  want  to  preserve   the  holistic  strategy,
monitoring a large number of  pairs will require a dedicated facility,
e.g. a  pair of 2  m robotic  telescopes in both  hemispheres equipped
with speckle imagers.







\begin{acknowledgments}

I thank  B.~Mason for extracting  historic data from the  WDS database
and for examination of the CHARA  data on B~868. Comments by R.~Matson
on  the paper  draft are  appreciated. This  work used  data from  the
Washington  Double Star  Catalog maintained  at USNO  and observations
made at  SOAR.  This work used  the SIMBAD service operated  by Centre
des   Donn\'ees   Stellaires   (Strasbourg,   France),   bibliographic
references from  the Astrophysics Data System  maintained by SAO/NASA.
This work  has made use of  data from the European  Space Agency (ESA)
mission Gaia (\url{https://www.cosmos.esa.int/gaia})  processed by the
Gaia   Data   Processing   and  Analysis   Consortium   (DPAC,   {\url
  https://www.cosmos.esa.int/web/gaia/dpac/consortium}).   Funding for
the DPAC has been provided by national institutions, in particular the
institutions participating in the Gaia Multilateral Agreement.

\end{acknowledgments} 


\facility{SOAR}


\bibliography{orb.bib}
\bibliographystyle{aasjournal}



\end{document}